\documentclass[reprint,amsmath,amssymb,aps,prb]{revtex4-2}

\usepackage{graphicx}
\usepackage{dcolumn}
\usepackage{bm}
\usepackage{bbold}
\usepackage[pdfusetitle,
            bookmarks=true,
            colorlinks,
            linkcolor=blue,
            citecolor=blue,
            urlcolor=blue
            ]{hyperref}
\usepackage{xcolor}
\usepackage{gensymb}
\usepackage{listings} 

\usepackage{physics}

\definecolor{codegreen}{rgb}{0,0.6,0}
\definecolor{codegray}{rgb}{0.5,0.5,0.5}
\definecolor{codepurple}{rgb}{0.58,0,0.82}
\definecolor{backcolour}{rgb}{0.95,0.95,0.92}

\lstdefinestyle{mystyle}{
    backgroundcolor=\color{backcolour},   
    commentstyle=\color{codegreen},
    keywordstyle=\color{magenta},
    numberstyle=\tiny\color{codegray},
    stringstyle=\color{codepurple},
    basicstyle=\ttfamily\footnotesize,
    breakatwhitespace=false,         
    breaklines=true,                 
    captionpos=b,                    
    keepspaces=true,                 
    numbers=left,                    
    numbersep=5pt,                  
    showspaces=false,                
    showstringspaces=false,
    showtabs=false,                  
    tabsize=2
}

\renewcommand{\O}{\mathcal{O}}

\newcommand\numberthis{\addtocounter{equation}{1}\tag{\theequation}}

\newcommand{\vc}[2]{\left(\!
    \begin{array}{c}
      #1 \\
      #2
    \end{array}
  \!\right)}

\newcommand{\eps}{\varepsilon}

\newcommand{\kk}{\mathbf{k}}

\newcommand{\qb}{\mathbf{q}}

\newcommand{\mfp}{\chi}

\newcommand{\ddelta}{\boldsymbol{\delta}}

\newcommand{\cchi}{\boldsymbol{\chi}}

\newcommand{\ii}{\mathbf{i}}

\newcommand{\bexpval}[1]{\big\langle#1\big\rangle}

\newcommand{\tfr}{\mathfrak{t}}
\newcommand{\bfr}{\mathfrak{b}}

\begin{document}

\title{Probing magnetism in moir{\'e} heterostructures with quantum twisting microscopes}

\newcommand{\TUM}{\affiliation{Technical University of Munich, TUM School of Natural Sciences, Physics Department, 85748 Garching, Germany}}
\newcommand{\MCQST}{\affiliation{Munich Center for Quantum Science and Technology (MCQST), Schellingstr. 4, 80799 M{\"u}nchen, Germany}}

\author{Fabian Pichler} \TUM \MCQST
\author{Wilhelm Kadow} \TUM \MCQST
\author{Clemens Kuhlenkamp} \TUM \MCQST
\author{Michael Knap} \TUM \MCQST

\date{\today}

\begin{abstract}
 Spin-ordered states close to metal-insulator transitions are poorly understood theoretically and challenging to probe in experiments. Here, we propose that the quantum twisting microscope, which provides direct access to the energy-momentum resolved spectrum of single-particle and collective excitations, can be used as a novel tool to distinguish between different types of magnetic order. To this end, we calculate the single-particle spectral function and the dynamical spin-structure factor for both a ferromagnetic and antiferromagnetic generalized Wigner crystal formed in fractionally filled moir{\'e} superlattices of transition metal dichalcogenide heterostructures. We demonstrate that magnetic order can be clearly identified in these response functions. Furthermore, we explore signatures of quantum phase transitions in the quantum twisting microscope response. We focus on the specific case of triangular moir{\'e} lattices at half filling that have been proposed to host a topological phase transition between a chiral spin liquid and a 120-degree ordered state. Our work demonstrates the potential for quantum twisting microscopes to characterize quantum magnetism in moir{\'e} heterostructures.
\end{abstract}

\maketitle

\section{Introduction}

Magnetic order emerging in proximity to metal-insulator transitions~\cite{imada_metal-insulator_1998, senthil_theory_2008} is challenging to study both theoretically and experimentally. Bilayers of transition metal dichalcogenides (TMDs)~\cite{manzeli_2d_2017} provide a new platform to realize such states, with high tunability and access to novel probes. TMD bilayers form moir\'e superlattices when the two layers are twisted by a small twist angle with respect to each other or when there is an intrinsic lattice mismatch in hetero bilayer structures. These moir\'e superlattices effectively simulate extended Hubbard models on a honeycomb or triangular lattice~\cite{wu_hubbard_2018, tang_simulation_2020, pan_quantum_2020}, which allows for the experimental realization of strongly correlated electron phases. Due to their large lattice constants on the order of $\sim 10$ nm, moir\'e materials allow for tuning through a wide range of fillings. At several fractional fillings, generalized Wigner crystals (GWC) have been observed~\cite{regan_mott_2020, shimazaki_strongly_2020, wang_correlated_2020, xu_correlated_2020, huang_correlated_2021, jin_stripe_2021, li_imaging_2021,  tang_evidence_2023}. These states break the discrete translation symmetry of the moir\'e lattice and form due to long-range interactions between electrons. Long-range interactions are relevant in moir\'e superlattices~\cite{morales-duran_non-local_2022} because the Wannier orbitals of the electrons are less localized, and the charge density is low~\cite{wu_hubbard_2018, wu_topological_2019, pan_band_2020, pan_quantum_2020}. While the charge order of the GWC states has been firmly established in experiments, their spin order remains an open question. This is due to competing contributions to the effective spin coupling, primarily consisting of antiferromagnetic superexchange and ferromagnetic direct exchange~\cite{roger_MultipleExchange_1984, Tanatar_GroundState_1989, Zhu_VariationalQuantum_1995, hu_competing_2021, morales-duran_non-local_2022, kim_InterstitialInducedFerromagnetism_2022, morales-duran_magnetism_2023}. Current methods to probe magnetic order are mainly based on magneto-optical methods measuring the magnetic susceptibility, suggesting a weak antiferromagnetic coupling at half filling~\cite{tang_simulation_2020, tang_evidence_2023}. Moreover, almost vanishing magnetic couplings in another sample have been found~\cite{ciorciaro_kinetic_2023}. These experimental findings are in stark contrast to theoretical predictions, which suggest large ferromagnetic interactions~\cite{hu_competing_2021, morales-duran_non-local_2022, morales-duran_magnetism_2023}.

To reliably determine the fate of spin order in GWCs, additional experimental input is required, which hinges on developing more direct experimental probes of quantum magnetism. One proposed method uses the Zeeman splitting in circularly polarized exciton Umklapp resonances to detect magnetic order~\cite{salvador_optical_2022, julku_exciton_2023}. Using such exciton Umklapp resonances, signatures of Wigner crystals in TMDs have been detected~\cite{shimazaki_optical_2021, smolenski_observation_2021}. Another approach to characterize the magnetic states is to use spin-polarized scanning tunneling spectroscopy or noise magnetometry~\cite{chatterjee_diagnosing_2019, feldmeier_local_2020, konig_tunneling_2020}. 

In this work, we propose that a Quantum Twisting Microscope (QTM)~\cite{Inbar_2023} can be used to distinguish between antiferromagnetic (AFM) and ferromagnetic (FM) spin order by measuring the energy and momentum resolved single-particle and collective-excitation spectrum. The QTM is designed for studying the dynamical response of two-dimensional moir\'e materials~\cite{Inbar_2023, peri_probing_2024}. Placing a layer of graphene on a capped pyramid on the edge of an atomic force microscope cantilever makes it possible to continuously twist two sheets of 2D materials with respect to each other, see Fig.~\ref{fig:1_v2}a). By applying a voltage between the top layer on the cantilever and the bottom sample layer and measuring the tunneling current as a function of the relative twist angle between the two layers, one obtains the momentum-resolved single-particle spectral function of the sample~\cite{Inbar_2023}. This is a form of elastic tunneling spectroscopy. In this work, we show that the expected spectral function for the GWC is qualitatively different for antiferromagnetic (AFM) and ferromagnetic (FM) interactions. An even more striking difference between the AFM and FM state can be seen in the dynamical spin-structure factor, which is accessible using inelastic tunneling spectroscopy~\cite{carrega_tunneling_2020, peri_probing_2024}. For that, the original setup of the QTM has to be modified to a three-layer structure, with graphene contacts on both sides of the sample, see Fig.~\ref{fig:2_v1}a). Applying a bias voltage between the graphene sheets, the spin response of the sample layer, which acts as a tunneling barrier, can be measured. Electrons tunneling from the top to bottom layer interact with low-energy excitations of the sample layer, thus measuring its collective excitation spectrum. Using both elastic and inelastic tunneling spectroscopy to directly probe the momentum-resolved excitation spectrum of two-dimensional materials, such as TMD heterostructures, provides new insight into the strongly correlated states of these systems, not accessible using conventional optical measurements.

This work is structured as follows: In Sec.~\ref{sec:elastic}, we review elastic tunneling spectroscopy,  and in Sec.~\ref{sec:MF_1},  we calculate the expected signal for both an AFM and FM GWC. In Sec.~\ref{sec:inelastic}, we review inelastic tunneling spectroscopy, and in Sec.~\ref{sec:coll_2}, we show the qualitative differences in the dynamical spin-structure factor between the AFM and FM GWC. In Sec.~\ref{sec:crit}, we propose that the QTM can also be used to probe quantum phase transitions, in particular focusing on a possible transition between a chiral spin liquid (CSL) and the ordered $120\degree$ state in the half-filled Hubbard model on a triangular lattice, where the onset of order manifests itself in a softening of a collective mode at the ordering wave vector. We conclude in Sec.~\ref{sec:conclusion} with an outlook. Technical details are deferred to the appendices.

\section{Single-particle spectral function from elastic tunneling} \label{sec:elastic}

\begin{figure}
\begin{center}
\includegraphics[width=\linewidth]{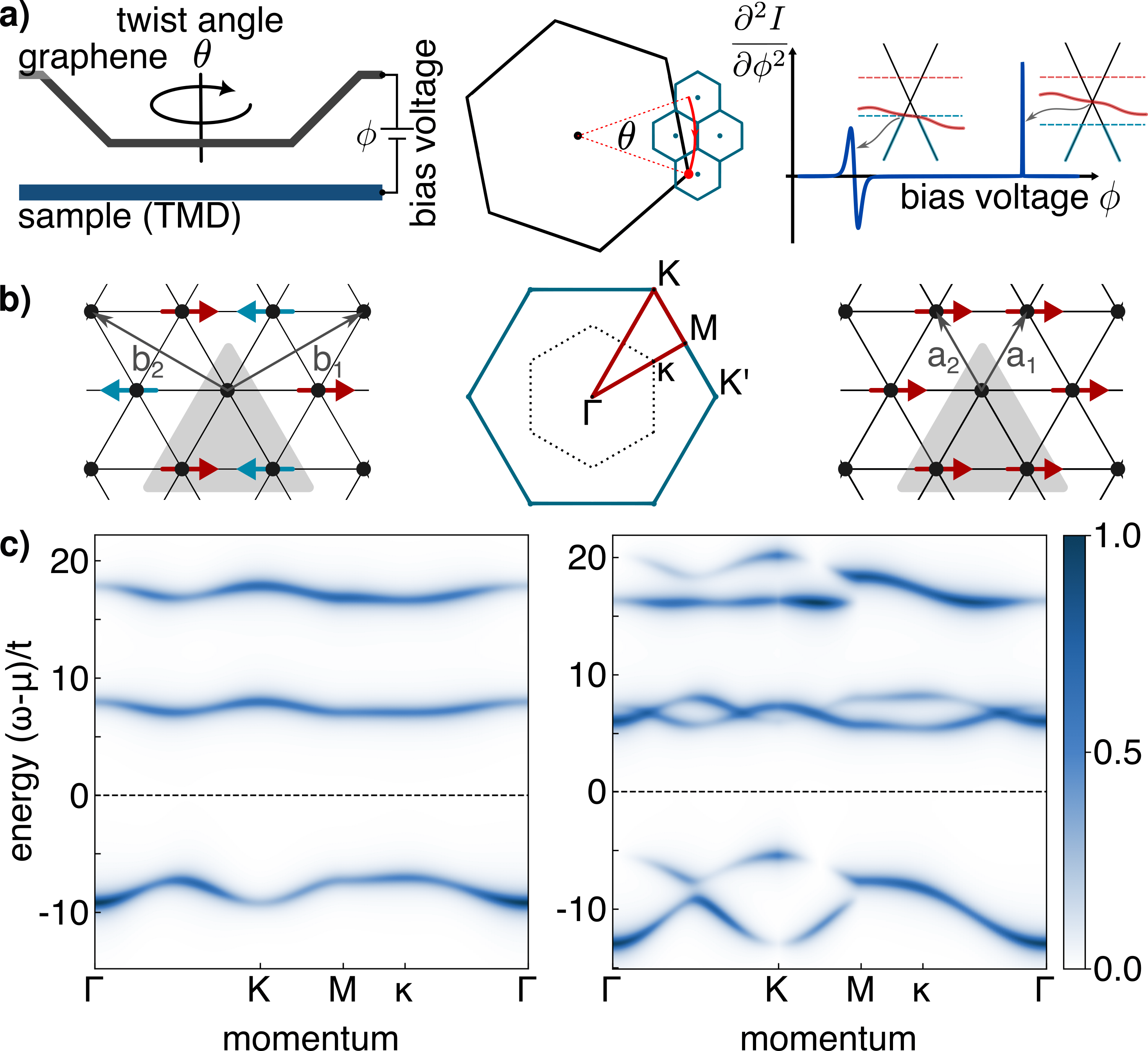}
\caption{\textbf{Single-particle spectral function for generalized Wigner crystals.} \textbf{a)} Left: Illustration of the QTM with a graphene layer on top of the sample layer. The top layer is twisted by an angle $\theta$ with respect to the bottom layer, and a voltage $\phi$ is applied between the two layers. Middle: By changing the twist angle $\theta$, the Dirac points of graphene, located at the corners of its Brillouin zone, trace a path through momentum space (shown in red). For moir\'e lattices, the Brillouin zone of the bottom layer (blue) is much smaller than the one of graphene (black), such that a small twist angle already cuts through the entire Brillouin zone of the sample. Right: Illustration of the second derivative of the tunneling current as a function of the applied voltage $\phi$ for fixed twist angle. There are two main features: a sharp temperature-independent peak at the Dirac matching condition, arising when the Dirac point of graphene touches an energy surface of the bottom layer, and a resonance peak at the onset condition when tunneling becomes energetically allowed. The tunneling current is proportional to the intersection length between the energy surfaces within the allowed energy interval set by the chemical potentials. \textbf{b)} Schematics of antiferromagnetic (left) and ferromagnetic (right) generalized Wigner crystal states. The chosen unit cell is highlighted in grey. The central panel shows the Brillouin zone of the sample (blue), including the reduced Brillouin zone (dashed) and the path used to plot the spectral functions. \textbf{c)} Spectral function $\mathcal{A}(\kk, \omega)$ for the antiferromagnetic (left, $U=25t$, $V=5t$, $X=0$) and ferromagnetic phase (right, $U=25t$, $V=5t$, $X=0.3t$). }
\label{fig:1_v2}
\end{center}
\end{figure}

The schematic setup for the QTM, realized experimentally in Ref.~\cite{Inbar_2023}, is shown in Fig.~\ref{fig:1_v2}a). The idea is similar to a conventional scanning tunneling microscope, but an extended sheet of graphene replaces the tip of the microscope. The tunneling current $I(\theta, \phi)$ between the graphene layer and the bottom layer, which one wants to examine, is measured as a function of the applied voltage $\phi$ and the relative twist angle $\theta$. To avoid hybridization between the graphene layer and the sample, an insulating barrier, such as hexagonal boron nitride, is placed inbetween. The formation of additional moir\'e structures  can be avoided by a large angle mismatch between the layers. Compared to a scanning tunneling microscope, spatial resolution is lost in this setting. Instead, the QTM has a momentum space resolution given by the linear size of the graphene layer $\delta k \sim L^{-1}$. The tunneling current is a convolution of the spectral functions $\mathcal{A_\ell(\kk, \omega)}$ of the top and bottom layer~\cite{Inbar_2023}

\begin{align*}
    I(\theta, \phi) = 4 \pi e |\Gamma_0|^2 \int &\dd \eps\; \big(f_T(\eps-\phi)-f_B(\eps)\big) \numberthis \label{eq:curr_1}\\ 
    \times&\displaystyle\sum_{\kk} \mathcal{A}_B(\kk, \eps) \mathcal{A}_T(\kk, \eps-\phi)
\end{align*}
where $f_\ell(\omega)$ is the Fermi-Dirac distribution, $\ell \in \{B, T\}$ is the layer index and $|\Gamma_0|^2$ is the elastic tunneling amplitude. 
A detailed derivation is presented in Appendix~\ref{sec:tunCurrs}. The dependence of the tunneling current on the twist angle $\theta$ is implicitly contained in the momentum of the top layer, which is rotated with respect to the bottom layer. One can effectively scan through the Brillouin zone of the bottom layer by varying the twist angle $\theta$. This is illustrated in the central panel of Fig.~\ref{fig:1_v2}a), where the Dirac point of the graphene layer traces a path through momentum space. Since the lattice constant for the moir\'e lattice $a_M \sim 10$ nm is much larger than the lattice constant of graphene $a_\text{gr} \sim 0.25$ nm, a small twist angle $\theta$ is already sufficient to cross the entire moir\'e Brillouin zone. Generically, the Brillouin zones of the top and bottom layer will be incommensurate with respect to each other, which implies that there is no guarantee that the path traced by the top layer's Dirac point will cross high symmetry points of interest in the bottom layer's Brillouin zone. Nevertheless, it is possible to explore several different cuts and approximately reach high symmetry points by varying the twist angle over a wide enough range.

We assume tunneling only occurs near the $K$ and $K'$ points of the top layer's  Brillouin zone, where we can describe graphene as a gas of massless Dirac fermions. To develop an intuitive understanding of the QTM, focus on one $K$ point with its corresponding Dirac cone. If we assume that the spectral functions define a sharp energy surface in momentum space, then the convolution in Eq.~\eqref{eq:curr_1} can be understood by a simple geometric picture: The tunneling current is proportional to the length of the intersection between the Dirac cone and the energy surfaces defined by the spectral function of the bottom layer. A given intersection only contributes to the current if it lies within the energy interval defined by the chemical potentials $\mu_\ell$, such that one of the energy bands is filled while the other is empty. Since we assume the filling of the sample layer to be constant, we fix its chemical potential $\mu_B$, while $\mu_T$ can be varied. This allows for measuring both the particle and hole part of the spectral function. Assuming $\mu_T < \mu_B$, one measures electrons tunneling from the bottom to the top layer, effectively determining the hole spectrum of the sample. The situation is reversed for $\mu_T > \mu_B$, where the particle spectrum is measured. In the geometric picture, the position of the Dirac point can be tuned vertically by changing the electrostatic bias potential $\phi$ between the layers and horizontally by varying the twist angle $\theta$. If the Dirac point lies exactly on an energy surface of the bottom layer, there will be no contribution to the current since the density of states at this point vanishes (in the geometric picture described above, the length of the intersection between the surfaces is zero). This condition, dubbed \emph{Dirac-matching condition}, leads to a sharp peak in the second derivative of the tunneling current with respect to $\phi$. Compared to the typical energy scales $\eps_B$ within a moir\'e lattice with lattice constant $a_M$, the graphene's Fermi velocity $v_F$ is very large $\hbar v_F /a_M \gg \eps_B$. In this limit, the Dirac cone forms a very sharp momentum-resolved tip, which scans the spectral function of the bottom layer. The latter is generically written as
\begin{equation}
    \mathcal{A}_B(\kk, \omega) = \displaystyle\sum_\alpha Z^\alpha_\kk \delta(\omega - \eps_B^\alpha(\kk)) + \text{continuum}
\end{equation}
with a band index $\alpha$ and spectral weight $Z^\alpha_\kk$. This leads to the tunneling current
\begin{equation}
    I(\qb(\theta), \phi) \propto \displaystyle\sum_\alpha Z_\qb^\alpha | \eps_B^\alpha (\qb) - \phi| \big[f_T(\eps^\alpha_B(\qb)-\phi) - f_B(\eps^\alpha_B(\qb))  \big]
\end{equation}
with $\qb(\theta) = \mathrm{R}(\theta)\mathbf{K} - \mathbf{g}$, where $\mathrm{R}(\theta)$ is a rotation matrix, and $\mathbf{g}$ is a reciprocal lattice vector, ensuring that $\qb$ lies within the first Brillouin zone of the bottom layer. Here, we have no longer explicitly written the contribution from the continuum part of the spectral function. 

More direct information is contained in the second derivative of the tunneling current
\begin{align*}
    \frac{\partial^2 I}{\partial \phi^2} &= \gamma_1 \displaystyle\sum_\alpha Z^\alpha_\qb \delta(\eps_B^\alpha(\qb) - \phi) \numberthis \\ &\times \big[f_T(\eps^\alpha_B(\qb)-\phi)-f_B(\eps^\alpha_B(\qb)) \big] \\
    & + \gamma_2 \displaystyle\sum_\alpha Z^\alpha_\qb \frac{\partial}{\partial \phi} \left( |\eps_B^\alpha(\qb) - \phi| \frac{\partial}{\partial \phi} f_T(\eps_B^\alpha(\qb) - \phi)\right) \label{eq:ETS}
\end{align*}
with some constants $\gamma_1 \sim \gamma_2$. In the rightmost panel of Fig.~\ref{fig:1_v2}a), a typical example of $\partial_\phi^2 I$ for fixed twist angle is sketched. There are two contributions: a distinct peak for $\eps_B^\alpha(\qb) =  \phi$ corresponding to the Dirac-matching condition discussed above and a second feature stemming from the $\phi$-dependence of the energy-interval in which tunneling is allowed. This second feature manifests as a maximum followed immediately by a minimum of the second derivative of the current at the point where tunneling first becomes allowed energetically. This resonance peak is dubbed the \emph{onset condition}~\cite{Inbar_2023}. Note that the strength of the peak of the Dirac-matching condition is based solely on density of states arguments and hence does not significantly depend on temperature. In contrast, the peaks associated with the onset condition get less pronounced with higher temperatures. Furthermore, by choosing the chemical potential of the top layer to be sufficiently large compared to the bias voltage $|\mu_T| \gg \phi$, one can neglect the contribution from the onset condition. In that case, $\partial^2_\phi I$ directly measures the spectral function of the bottom layer. 

\section{Probing single-particle excitations of generalized Wigner crystals} \label{sec:MF_1}

The QTM can be used to identify different magnetic correlations. We explicitly calculate the single-particle spectral function for GWCs with antiferromagnetic and ferromagnetic spin order and show that the different states have rather distinct signatures. 
We describe the TMD heterobilayer moir\'e superlattice with an extended Hubbard model~\cite{morales-duran_non-local_2022, wu_hubbard_2018}
\begin{align*}
    H &= - t \displaystyle\sum_{\expval{i, j} \sigma} (c_{\sigma i}^\dagger c_{\sigma j} + \text{h.c.}) + U \displaystyle\sum_{i} n_{\uparrow i} n_{\downarrow i} +  \\
    &\quad \quad +V \displaystyle\sum_{\substack{\expval{i j} \\ \sigma, \sigma'}} n_{\sigma i} n_{\sigma' j} + X \displaystyle\sum_{\substack{\expval{i j} \\ \sigma \sigma'}} c_{\sigma i}^\dagger c_{\sigma' j}^\dagger c_{\sigma' i} c_{\sigma j} \numberthis \\
     &= H_t + H_u + H_v + H_x \label{eq:extHubbard}
\end{align*}
with the number operator $n_{\sigma, i} = c_{\sigma, i}^\dagger c_{\sigma, i}$. Both the on-site interaction $U$ and the nearest-neighbor interaction strength $V$ are repulsive. Additionally, an intersite-exchange term $X$ is included, which gives a negative contribution to the energy when two neighboring spins are aligned and a positive contribution when misaligned, thus favoring ferromagnetism. We perform a mean-field decoupling of the interaction terms of the Hamiltonian. To this end, we introduce the following notation for the order parameter $\mfp_{\sigma \sigma'}^{i j} := \bexpval{c_{\sigma i}^\dagger c_{\sigma' j}}$. The resulting mean-field Hamiltonian reads:
\begin{align*}
    H_u &\simeq U \displaystyle\sum_{i, \sigma} (\mfp_{\sigma \sigma}^{i i} c_{\bar{\sigma} i}^\dagger c_{\bar{\sigma} i} - \mfp_{\sigma \bar{\sigma}}^{i i} c_{\bar{\sigma} i}^\dagger c_{\sigma i}) \\
    &\quad - U \displaystyle\sum_{i}(\mfp_{\uparrow \uparrow}^{i i} \mfp_{\downarrow \downarrow}^{i i} - \mfp_{\uparrow \downarrow}^{i i} \mfp_{\downarrow \uparrow}^{i i}) \numberthis\\
    H_v &\simeq V \displaystyle\sum_{\expval{ij}}\displaystyle\sum_{\sigma \sigma'} \big( \mfp_{\sigma \sigma}^{i i} c_{\sigma' j}^\dagger c_{\sigma' j} + \mfp_{\sigma' \sigma'}^{j j} c_{\sigma i}^\dagger c_{\sigma i} - \mfp_{\sigma' \sigma}^{j i} c_{\sigma i}^\dagger c_{\sigma' j} \\ 
    &\quad - \mfp_{\sigma  \sigma'}^{i j} c_{\sigma' j}^\dagger c_{\sigma i} - \mfp_{\sigma \sigma}^{i i} \mfp_{\sigma' \sigma'}^{j j} + \mfp_{\sigma' \sigma}^{j i} \mfp_{\sigma \sigma'}^{i j}\big) \numberthis\\
    H_x &\simeq X\displaystyle\sum_{\expval{ij}} \displaystyle\sum_{\sigma \sigma'} \big( \mfp_{\sigma \sigma'}^{i i} c_{\sigma' j}^\dagger c_{\sigma j} + \mfp_{\sigma' \sigma}^{j j} c_{\sigma i}^\dagger c_{\sigma' i} - \mfp_{\sigma \sigma}^{i j} c_{\sigma' j}^\dagger c_{\sigma' i} \\ 
    &\quad - \mfp_{\sigma' \sigma'}^{j i} c_{\sigma i}^\dagger c_{\sigma j} - \mfp_{\sigma' \sigma}^{j j} \mfp_{\sigma \sigma'}^{i i} + \mfp_{\sigma \sigma}^{i j} \mfp_{\sigma' \sigma'}^{j i}\big). \numberthis
\end{align*}
In the following, we focus on a filling of $\nu=2/3$ electrons per site, where the formation of GWCs is expected~\cite{watanabe_charge_2005, tocchio_phase_2014, padhi_generalized_2021, morales-duran_magnetism_2023}. We introduce a three-site unit cell, which is sufficient to accommodate the expected spin and charge order at this filling. The corresponding fermionic creation (annihilation) operator for site $i$ on sublattice $a\in \{A, B, C\}$ and spin $\sigma \in \{\uparrow, \downarrow\}$ is denoted by $c_{a\sigma i}^\dagger$ ($c_{a \sigma i}$). By translation invariance, the mean-field parameters only depend on the relative distance between sites $i$ and $j$. Consequently, we can rewrite them as matrices in sublattice space: $\mfp_{\sigma \sigma'}^{a b} := \bexpval{c_{a \sigma i}^\dagger c_{b \sigma' i}}$. Next, we perform a Fourier transformation to momentum space
\begin{equation}
    c_{a \sigma i} = \frac{1}{\sqrt{N}} \displaystyle\sum_\kk e^{i \kk \cdot \mathbf{x}_i} c_{a \sigma \kk} ,
\end{equation}
where $N$ is the number of unit cells, and the momentum sum is over the reduced Brillouin zone obtained by increasing the unit cell to contain three sites. We bring the mean-field Hamiltonian into bilinear form
\begin{equation}
    H = \displaystyle\sum_\kk \Psi_\kk^\dagger h_\kk \Psi_\kk  + E_0  \label{eq:bilHam}
\end{equation}
with some constant energy $E_0$, the sublattice spinor $\Psi_\kk = (c_{A \uparrow \kk}, c_{A \downarrow \kk}, c_{B \uparrow \kk}, c_{B \downarrow \kk}, c_{C \uparrow \kk}, c_{C\downarrow \kk})^T$ and a $6\times6$ matrix $h_\kk$. The matrix $h_\kk$ depends on the mean-field parameters $\mfp_{\sigma \sigma'}^{ab}$, which must be self-consistently determined. To find a self-consistent mean-field solution, we iteratively diagonalize the Hamiltonian of Eq.~\eqref{eq:bilHam} and use the resulting ground state to compute $\mfp_{\sigma \sigma'}^{ab}$. The explicit form for the matrix $h_\kk$ and the energy shift $E_0$ is given in Appendix~\ref{sec:MFT}. The chemical potential is determined implicitly by fixing the electron density $\nu$, given by the number of electrons per site.

Our mean-field calculations for $\nu=2/3$ are in good agreement with previous mean-field~\cite{watanabe_charge_2005, tocchio_phase_2014, zang_hartree-fock_2021, qin_effect_2022} and exact-diagonalization studies~\cite{morales-duran_magnetism_2023}. We find a disordered metallic phase for small interactions $U$ and $V$ and small intersite direct exchange $X$. By increasing the onsite-interaction $U$, the ground state undergoes a crystallization transition and forms an antiferromagnetic spin state, where one sublattice is completely empty. Hence, the electrons arrange themselves in an effective hexagonal lattice as shown in the left panel of Fig.~\ref{fig:1_v2}b). Remarkably, the phase survives on a mean-field level down to $V=0$, where one would naively no longer expect the stabilization of hexagonal charge order. This configuration is favored due to an interplay between spin and charge ordering. By crystallizing in a honeycomb lattice, the electrons can access an unfrustrated antiferromagnetic spin configuration, lowering their energy. By increasing the intersite direct exchange term $X \gtrsim 0.2t-0.4t$, the effective nearest-neighbor spin coupling changes sign~\cite{morales-duran_magnetism_2023}, and we find a ferromagnetic ground state. The corresponding real-space spin configuration is shown in the right panel of Fig.~\ref{fig:1_v2}b).

Excitations on top of the antiferromagnetic and ferromagnetic phases differ strongly, leading to distinct experimental signatures. We present theoretical predictions for the QTM response of the different GWCs and how these signatures can be used to distinguish between the different phases. First, we calculate the single-particle spectral function
\begin{equation}
    \mathcal{A}(\kk, \omega) = - \frac{1}{\pi} \Im \mathcal{G}(\kk, \omega) \label{eq:specfunc}
\end{equation}
with the retarded Green's function  defined as
\begin{equation}
    \mathcal{G}(\kk,  \omega) = - i \displaystyle\sum_\sigma \displaystyle\int_0^\infty \dd t \; e^{i \omega t} \bexpval{\{c_{\sigma \kk}(t), c^\dagger_{\sigma \kk}(0)\}} \label{eq:GF}
\end{equation}
and the expectation value taken with respect to the ground state. Since our main goal is to show how the QTM can distinguish different GWCs, we are only interested in a qualitative description. 
To characterize the main features of the spectral function, we focus on the bare mean-field response~\cite{kadow_hole_2022}. Computational details are given in Appendix~\ref{sec:MFT}. In Fig.~\ref{fig:1_v2}c), we present representative results for the spectral functions in the AFM (left panel) and FM (right panel) phases. 

In the antiferromagnetic GWC phase, we observe one hole band and two independent particle bands. The two distinct particle bands can be understood intuitively: there are two different ways to add an electron onto the AFM ground state, either by putting it on an empty site or by putting it on an occupied site. The gap and the bandwidth can be calculated using second-order perturbation theory. The gap between the two particle bands is approximately $\Delta E = U - 3V$ while the bandwidth is of order $t^{(\text{eff})}_1 \sim t^{\text{eff}}_2 \sim \O(t^2/U)$, assuming $U \gg V, t$. Detailed expressions for $t^{\text{eff}}_{1/2}$ are given in Appendix~\ref{sec:eff_hop}. Similar arguments can be made for the hole band, leading to a gap of $\Delta E = 3V$ between the hole band and the lower particle band. The bandwidth for the hole band is of the same order as for the particle bands. 

The spectral function in the ferromagnetic GWC is qualitatively different. The most significant difference is that the number of mean-field bands doubles. This can be understood as a Zeeman splitting due to the non-vanishing overall magnetization. The individual bands can again be interpreted as a particle or hole moving in a mean-field background: the two topmost particle bands result from putting a particle on an already occupied site. It can freely hop on an effective hexagonal lattice, leading to a graphene-like dispersion. The closing of the gap between the two bands at the corners $\kappa$ of the reduced Brillouin zone is not visible in Fig.~\ref{fig:1_v2}c), as only one band has spectral weight at $\kappa$. The bandwidth of these two bands is of order $\O(t)$. The other two particle bands in the FM can be understood by putting a particle on an empty site. There are two possibilities: either the spin of the additional particle is aligned with its neighbors, or it is misaligned. This results in a separation of $\Delta E = 6X$, which is of a similar order as the bandwidth $\O(t^2/U)$, leading to several crossings. Also, the hole spectrum in the ferromagnet is graphene-like since a single hole on top of the ground state can freely hop on the effective hexagonal lattice. In contrast to the AFM-hole band, the ferromagnetic one has a bandwidth of order $\O(t)$.   
\\

In real materials, additional features will be present in the spectral function, which cannot be captured by mean-field theory. Nevertheless, we expect that the most prominent bands are the ones discussed above. 
It is possible that due to inhomogeneities of the sample, different patches of the sample are in different phases. As long as these puddles have the same size or are larger than the probing graphene sheet, which has a linear dimension of about~$200$~nm~\cite{Inbar_2023}, the QTM can distinguish between different phases even within the same sample. In that way, the QTM is ideally suited to probe the physics of large but spatially localized phases in the sample. This contrasts conventional transport measurements, where such inhomogeneities are more challenging to resolve.

Another possible challenge is the formation of mirror charges in the graphene layer due to charge order in the sample. This is relevant since we assume that the chemical potential in the graphene sheet is tuned away from charge neutrality. Well-defined Dirac cones are essential for the QTM to measure energy and momentum-resolved spectral functions. Consequently, we examine the influence of mirror charges on the Dirac cones. Since the expected charge order is at the moir\'e scale $a_M \gg a_\text{gr}$, we can take the long-wavelength limit to describe the graphene layer. The presence of mirror charges leads to an additional potential term $V(\mathbf{x})$ in the Hamiltonian:
\begin{equation}
    H =  H_0 + V(\mathbf{x}) = \hbar v_F (k_x \sigma^x + k_y \sigma^y) + V(\mathbf{x}).
\end{equation}
Due to the large Fermi velocity $v_F \approx 10^6$ m/s of graphene, the potential term can be treated as a small perturbation to the kinetic term. Since the potential varies spatially on the moir\'e scale $a_M$, which is much larger than the distance between the two sublattices of graphene, both sublattices feel, to leading order, the same potential. Any perturbation to $H_0$ proportional to the identity in sublattice space cannot gap out the Dirac cones of graphene. Hence, we expect the Dirac cones to remain intact to first approximation. There can be subleading terms not proportional to the identity, which stem from the small difference in potential between the two sublattices. Any such perturbation scales with $\lambda = a_\text{gr}/a_M \sim 10^{-2}$. To analyze this effect, we consider a generic periodic potential of the form
\begin{equation}
    V(\mathbf{x}) = \displaystyle\sum_\mathbf{G} \lambda V_\mathbf{G} e^{i \mathbf{G} \cdot \mathbf{r}} \sigma^z.
\end{equation}
Assuming that there is no constant energetic offset between the sublattices, the gap at the Dirac point scales as $\Delta \sim (\lambda V)^3$ to leading order in perturbation theory. Note that the cubic scaling is a consequence of the particle-hole symmetric dispersion of graphene and contrasts the generic quadratic scaling obtained for a free dispersion of the form $k^2/2m$. 
We conclude that the gap at the Dirac points is strongly suppressed by a factor of $\lambda^3$, which should ensure the gap to be smaller than the probed energy scale.

We have shown that by directly measuring these bands using the QTM, different magnetic states in GWCs can be distinguished. In particular, the spin degenerate single-particle bands in the AFM split due to a non-vanishing overall magnetization in the FM state. This has the potential to resolve an outstanding puzzle in the field: Simple theoretical estimates yield large values for the exchange coupling $X$~\cite{hu_competing_2021, morales-duran_magnetism_2023}. In stark contrast, susceptibility measurements do not reveal the expected strong ferromagnetism~\cite{tang_simulation_2020, tang_evidence_2023, ciorciaro_kinetic_2023}. The QTM response may provide useful insight into this puzzle.

\section{Collective excitation spectrum from inelastic tunneling} \label{sec:inelastic}

\begin{figure}
\begin{center}
\includegraphics[width=\linewidth]{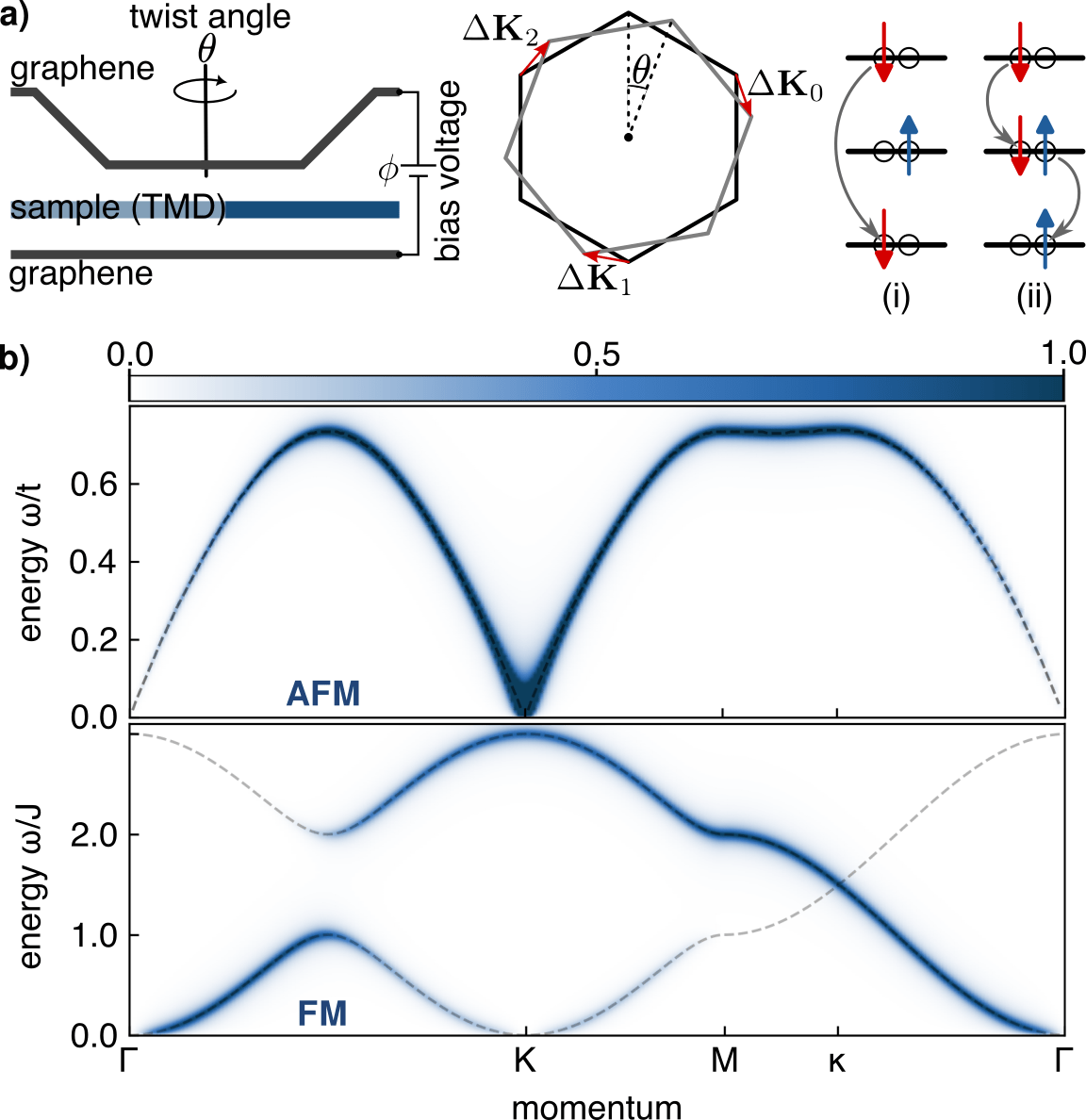}
\caption{\textbf{Dynamical spin-structure factor for generalized Wigner crystals.} \textbf{a)} Left: QTM with an intermediate sample layer sandwiched between two layers of graphene. The top layer is twisted by an angle $\theta$ with respect to the bottom layer, and a voltage $\phi$ is applied between the two graphene layers. Middle: Brillouin zones of the top and bottom layer graphene twisted by an angle $\theta$. The measured dynamical spin-structure factor is a function of $\Delta \mathbf{K}_n$. Right: Different processes contributing to the total tunneling current: $(i)$ elastic tunneling from top to bottom layer; $(ii)$ inelastic tunneling where the electron virtually tunnels to the intermediate layer, effectively interacting with spin excitations of that layer. \textbf{b)} Dynamical spin-structure factor $\mathcal{S(\kk, \omega)}$ of the antiferromagnetic (top) and ferromagnetic phase (bottom). The dotted lines correspond to the spin-wave dispersion from linear spin-wave theory.}
\label{fig:2_v1}
\end{center}
\end{figure}

Inelastic tunneling spectra can be measured with a QTM by sandwiching the sample in between two graphene layers~\cite{carrega_tunneling_2020, peri_probing_2024}, as illustrated in Fig.~\ref{fig:2_v1}a). Momentum resolution is again obtained by twisting the top layer by an angle $\theta$. When a bias voltage $\phi$ is applied between the two graphene layers, electrons can tunnel directly from the top to the bottom layer. This results in an elastic contribution to the tunneling current similar to the one discussed in Sec.~\ref{sec:elastic}. In addition, the electron can also scatter off the intermediate layer, which results in an inelastic contribution to the current. Through these inelastic tunneling processes, one can probe the collective excitation spectrum of the intermediate layer. Heuristically, the electron interacts with the low-energy excitations of the sample layer, allowing for contributions to the current in regimes where elastic tunneling is forbidden by energy-momentum conservation. Momentum resolution is again obtained by twisting the two layers of graphene with respect to each other. If the intermediate layer hosts spin excitations, which couple to the tunneling electrons with a coupling $J$, one will measure the dynamical spin-structure factor
\begin{equation}
    \mathcal{S}(\qb, i\omega_n) =  -\frac{1}{\pi} \Im \displaystyle\sum_\mu \displaystyle\int_0^\beta \dd \tau\; e^{i \omega_n \tau} \expval{\mathcal{T} s_\qb^\mu (\tau) s_{-\qb}^\mu (0)} \label{eq:spinspin}
\end{equation}
where $s^\mu_\qb(\tau)$ is the Fourier transformed local magnet moment in the intermediate layer and $\mathcal{T}$ is the imaginary time ordering operator. 
By choosing the chemical potentials of top and bottom layer to be sufficiently large $\mu_T = \mu_B \equiv \mu \gg \phi$ and by using again that the Fermi velocity of graphene is very large $\hbar v_F a_\text{gr} \gg \mu \theta $, one can write the contribution of the inelastic tunneling current as~\cite{peri_probing_2024}
\begin{equation}
    I^{(2)}(\theta, \phi) = \gamma \displaystyle\sum_{n=0}^2 \displaystyle\int_0^\phi \dd \omega\; (\phi -\omega) \mathcal{S}(\Delta\mathbf{K}_n, \omega) \label{eq:inelSIM}
\end{equation}
where $\Delta \mathbf{K}_n = (\mathrm{R}(\theta)-\mathbb{1})\mathbf{K}_n$ is the difference between the K points of the top and bottom layer's Brillouin zone and $\gamma$ is some constant. The index $n$ indicates the three different $K$ points, related by a reciprocal lattice vector, as shown in the central panel of Fig.~\ref{fig:2_v1}a). Note that there are additional contributions from tunneling around the $K'$ valley, which are related to the ones from the $K$ valley by time-reversal symmetry. Concretely, there are additional contributions in Eq.~\eqref{eq:inelSIM} with $\Delta \mathbf{K}_n$ replaced by $-\Delta \mathbf{K}_n$~\cite{peri_probing_2024}. The second derivative of the current directly measures the dynamical spin-structure factor 
\begin{equation}
    \frac{\partial^2 I^{(2)}}{\partial \phi^2} = \gamma\displaystyle\sum_{n=0}^2 \mathcal{S}(\Delta\mathbf{K}_n, \phi). \label{eq:ssf}
\end{equation}
While tunneling through the intermediate layer, the electrons may not only couple to spin excitations but also to density excitations. If $t$ is the hopping term from the top layer to the intermediate layer and $U$ the on-site interaction cost in the intermediate layer, then the effective spin coupling will be of order $J \sim t^2/U$ while the coupling to the density excitations will be of order $U$. In the insulating phase, $U/t \gg 1$ is large, and therefore, spin and density excitations lie at very different energy scales. Hence, they can be probed independently. For the contribution to the current obtained by coupling to density excitations, the result is the same as in Eq.~\eqref{eq:ssf}, but with the spin structure factor replaced by the density response
\begin{equation}
    \mathcal{S}_{\text{den}}(\qb, i\omega_n) =  -\frac{1}{\pi} \Im \displaystyle\int_0^\beta \dd \tau\; e^{i \omega_n \tau} \expval{\mathcal{T} \rho_\qb (\tau) \rho_{-\qb} (0)}. \label{eq:denden}
\end{equation}
In the kinematic region where elastic tunneling is allowed, its contribution dominates over the inelastic tunneling current, which is suppressed by a relative factor of $|J/\Phi|^2$. Here, $\Phi$ is the strength of the tunneling barrier between the top and bottom layers. Consequently, the QTM can only probe the dynamical structure factor in the absence of elastic tunneling. The condition for elastic tunneling, given by~\cite{peri_probing_2024},
\begin{equation}
    \phi < \hbar v_F |\Delta \mathbf{K}_n| < \phi + 2 \mu \label{eq:condIETS}
\end{equation}
can be tuned in an experiment by changing the chemical potential $\mu$ of the two graphene sheets.

\section{Collective Magnetic Excitations of generalized Wigner crystals} \label{sec:coll_2}
We have already discussed in Sec.~\ref{sec:MF_1} how the QTM can distinguish between GWCs with ferro- and antiferromagnetic spin order by measuring the single-particle spectral function. An even more decisive difference between the two states can be seen in the spin-spin response, accessible through inelastic tunneling as described in Sec.~\ref{sec:inelastic}. Collective modes can be captured by studying fluctuations around the mean-field state. Therefore we employ an RPA calculation~\cite{singh_quantum_1990, singh_finite-u_2005, knolle_theory_2010, willsher_magnetic_2023}, see Appendix~\ref{sec:RPA}.

In the top panel of Fig.~\ref{fig:2_v1}b), a representative dynamical spin-structure factor as defined in Eq.~\eqref{eq:spinspin} is shown for an antiferromagnetic GWC. We note that on top of the gapless spin-wave modes, with bandwidth $\O(J)$, there is a gapped particle-hole continuum at an energy scale $U \gg J$. The spin-wave modes become gapless at the $K$ and $K'$ points of the Brillouin zone, which correspond to reciprocal lattice vectors of the extended lattice with a three-site unit cell. We observe only a single mode, which is twofold degenerate. As expected for an antiferromagnet, the spin-wave dispersion is linear in the long-wavelength limit. Deep within the AFM phase ($U \gg V, t$), the spin-wave modes can be described using an antiferromagnetic Heisenberg model on a hexagonal lattice and applying a Holstein-Primakoff transformation. The resulting spin-wave dispersion is given  by
\begin{equation}
    \eps_\kk = \frac{3 J}{2} \sqrt{1- |\gamma_\kk|^2} \label{eq:lin_sw}
\end{equation}
with $\gamma_\kk = ( 1+ e^{i \kk \cdot \mathbf{b}_1} + e^{i \kk \cdot (\mathbf{b}_1+\mathbf{b}_2)})/3$. The effective nearest-neighbor spin coupling 
\begin{equation}
    J = \frac{4 t^2}{U - V} - 2 X \label{eq:spincoupl}
\end{equation}
contains the usual superexchange term and contributions from direct exchange processes $X$~\cite{morales-duran_magnetism_2023}. Note that because the underlying lattice is triangular, there are also contributions to the spin coupling from spin-flip processes where an electron virtually hops to an empty site. Both the RPA formalism used to compute the spin waves in the top panel of Fig.~\ref{fig:2_v1}b) and the linear spin-wave theory, which yields Eq.~\eqref{eq:lin_sw}, do not take interactions between the collective modes into account. Such interactions would lead to a finite lifetime and renormalization of the energy~\cite{mourigal_dynamical_2013}. Moreover, since the spin order is collinear, there are no cubic interaction terms for the spin waves, which is why we do not expect interaction effects between spin waves to influence their dispersion significantly~\cite{chernyshev_magnon_2006, chernyshev_spin_2009}. 

In agreement with our mean-field calculation, it is clear from Eq.~\eqref{eq:spincoupl} that a large intersite direct exchange term $X$ leads to a ferromagnetic spin coupling. A representative dynamical spin-structure factor for a ferromagnetic GWC is shown in the bottom panel of Fig.~\ref{fig:2_v1}b). It was calculated using linear spin-wave theory, see Appendix~\ref{sec:HPdisp}. The magnon spectrum can again be described using a Holstein-Primakoff transformation~\cite{pershoguba_dirac_2018}, yielding
\begin{equation}
    \eps_\kk^{\pm} = \frac{3 J}{2}(1\pm |\gamma_\kk|). \label{eq:magnonFM}
\end{equation}
 The lower magnon branch is gapless at the $\Gamma$, $K$, and $K'$ points of the Brillouin zone. In contrast to the antiferromagnetic case, the long wavelength behavior is quadratic instead of linear. Furthermore, the two modes are no longer degenerate. Instead, we have a second band, which is gapped at zero momentum. The upper and lower magnon bands touch at the corners $\kappa$ of the reduced Brillouin zone with linear dispersion, forming bosonic Dirac cones. We again note that magnon-magnon interactions renormalize the energy of the two bands and introduce a finite lifetime to the magnons~\cite{pershoguba_dirac_2018}. Still, the qualitative differences to the AFM case will remain. 

 Since the effective hexagonal lattice of the GWC requires a two-site unit cell independent of the spin order, there is no unit-cell doubling when going from the ferromagnetic to the antiferromagnetic phase. This is different at half-filling, with $\nu=1$ electron per site. Here, the ferromagnetic state does not require an extended unit cell, while the competing antiferromagnetic $120\degree$ state requires a three-site unit cell. The QTM can also be applied to the case of half-filling to determine the dominant spin coupling. In that case, the larger unit cell of the AFM leads to a smaller periodicity in momentum space of both the single-particle spectral function and the dynamical spin-structure factor. This is an additional qualitative difference between the two types of spin-order, which is present at $\nu=1$ and $\nu=1/3$, which both stabilize triangular charge order, but not at $\nu=2/3$, which stabilizes honeycomb charge order.

\section{Characterizing Quantum Criticality}\label{sec:crit}

\begin{figure}
\begin{center}
\includegraphics[width=\linewidth]{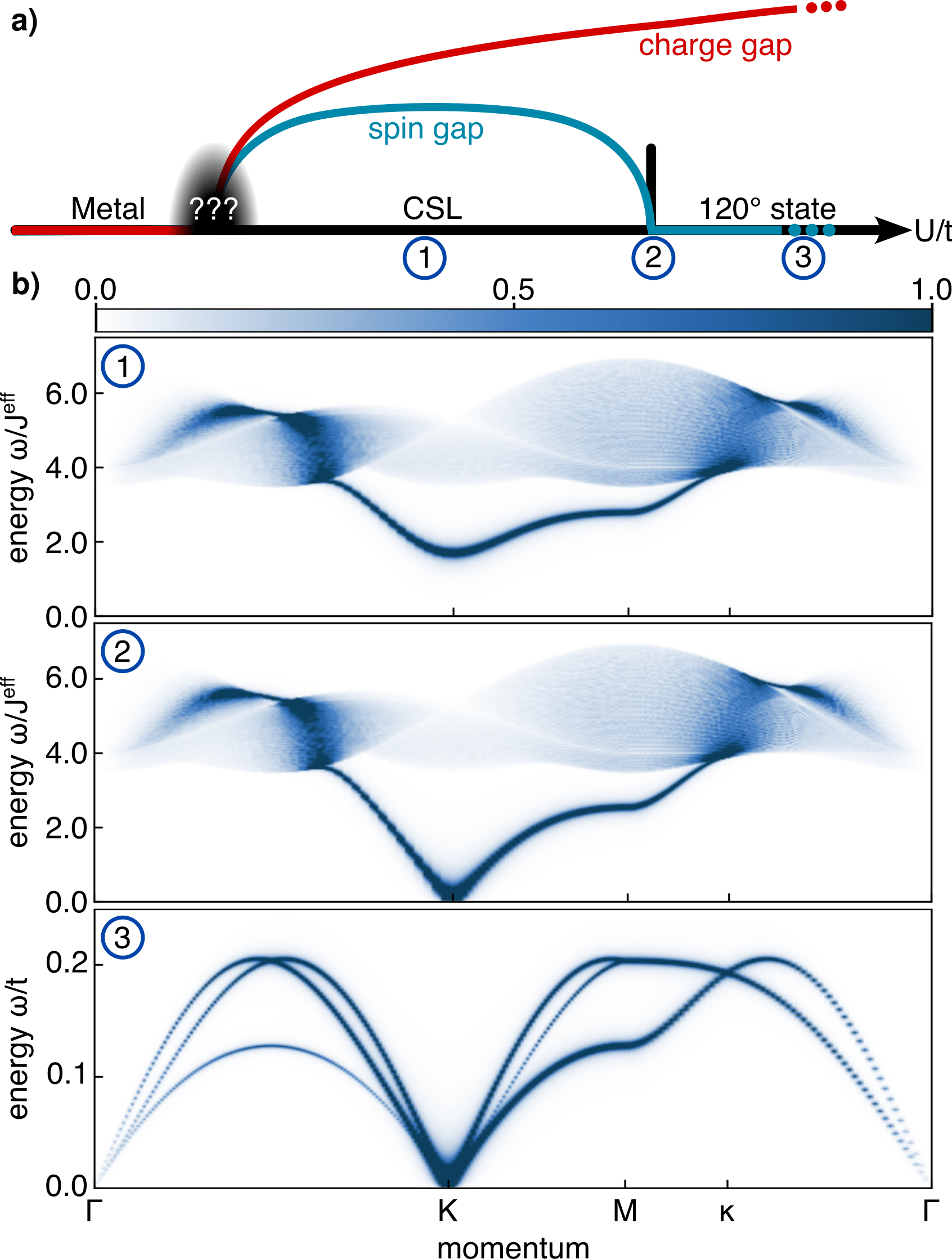}
\caption{\textbf{Spin response across a quantum phase transition.} \textbf{a)} Proposed zero temperature phase diagram of the half-filled triangular Hubbard model~\cite{szasz_chiral_2020, cookmeyer_four-spin_2021}. The charge and spin gap are schematically drawn as a function of interaction strength $U/t$. \textbf{b)} Dynamical spin-structure factor $\mathcal{S}(\kk, \omega)$ in the chiral spin liquid (CSL) phase (1), at the critical point (2) and in the ordered $120\degree$ phase (3). The roton in the CSL becomes soft at the $K$ point of the Brillouin zone at the critical point, indicating a transition to an ordered state.}
\label{fig:3_v1}
\end{center}
\end{figure}

So far, we have characterized the QTM response of spin orders in generalized Wigner crystals. Here, we discuss the QTM response in the vicinity of quantum phase transitions and demonstrate its potential as a new tool to explore critical phenomena in two-dimensional materials. As a concrete example, we focus on the half-filled Hubbard model on a triangular lattice. For simplicity, we assume $V=X=0$. For large $U/t$, the ground state is the $120\degree$-ordered state, while for small interaction strengths, we find a metallic phase. In between an intermediate Kalmeyer-Laughlin type chiral spin liquid (CSL) phase has been predicted~\cite{kalmeyer_equivalence_1987, szasz_chiral_2020, cookmeyer_four-spin_2021, chen_quantum_2022, kadow_hole_2022, kuhlenkamp_tunable_top_2022, motruk_KagomeChiral_2023a}.  

The charge and spin gap in the three phases are schematically shown in Fig.~\ref{fig:3_v1}a). The spin gap closes at the transition from the CSL to the $120\degree$ state. At the transition from the CSL to the metal, both the spin and charge gap must close. While the former is directly observable through inelastic tunneling spectroscopy, the closing of the charge gap can be measured using elastic tunneling spectroscopy with a QTM, where the single-particle spectral function is obtained. 
On the metallic side of the transition, one should be able to directly observe the vanishing of the quasiparticle residue $Z \sim |U - U_c|^{2 \beta}$, as predicted for a second order Mott transition~\cite{senthil_theory_2008}, with a critical exponent $\beta$. Furthermore, the zero sound velocity is expected to diverge. This is, in principle, also accessible through inelastic tunneling spectroscopy by measuring the density response in the metal. 
Here, we consider only one possible scenario for a transition between the CSL and the $120\degree$ state, where a continuous softening of a sharp roton mode at the $K$ point is observed, see Fig.~\ref{fig:3_v1}b). We emphasize that other transitions are conceivable, e.g., via an intermediate  Dirac spin liquid~\cite{song_unifying_2019}. Crucially, we argue that the QTM response would provide important information on the phases and their transitions.

We calculate the dynamical spin-structure factor of Eq.~\eqref{eq:spinspin}, which is experimentally accessible through inelastic tunneling spectroscopy in both the CSL and $120\degree$ phase. We employ the RPA formalism discussed in Appendix~\ref{sec:RPA} in both cases. The resulting spin-wave dispersion deep within the $120\degree$ phase is shown in panel (3) of Fig.~\ref{fig:3_v1}b). Due to the three-sublattice structure of the ordered state, there are three distinct modes, which become soft at the $\Gamma$, $K$, and $K'$ points of the Brillouin zone. Far from the transition, the three modes can again be understood using linear spin-wave theory. A Holstein-Primakoff transformation yields the following dispersion
\begin{equation}
    \eps_\kk = \frac{3 J}{2} \sqrt{(1+\gamma_\kk)(1+2\gamma_\kk)}
\end{equation}
where $\gamma_\kk = 1/6 \sum_{\ddelta=1}^6 e^{i \kk \cdot \ddelta}$ is a geometric factor resulting from a sum over all nearest-neighbors $\ddelta$ of the triangular lattice. This dispersion corresponds to a longitudinal mode in the long-wavelength limit~\cite{singh_finite-u_2005}. The remaining two modes, which are transversal for small momenta, are obtained from $\eps_{\kk + \mathbf{K}}$ and $\eps_{\kk + \mathbf{K}'}$. Note that along the high-symmetry line connecting the $\Gamma$ to the $M$-point in the Brillouin zone, two of the three modes become degenerate. Close to the transition, one of the three modes develops a roton-like minimum around the $M$-point (not shown), which becomes soft at a critical value of $U_{c}/t \simeq 6.8$~\cite{singh_finite-u_2005, willsher_magnetic_2023}. This indicates an instability of the ordered $120\degree$ phase, driven by out-of-plane fluctuations. 

In the CSL phase, we also calculate the dynamical spin-structure factor. We describe the CSL by the Hamiltonian
\begin{equation}
    H = - \frac{J}{2} \displaystyle\sum_{\substack{\expval{i j} \\ \alpha \beta}} f_{i \alpha}^\dagger f_{j \beta}^\dagger f_{i \beta} f_{j \alpha} \label{eq:HeisenHam}
\end{equation}
which is up to a constant simply the antiferromagnetic Heisenberg Hamiltonian expressed in terms of Abrikosov fermions $f_{i \sigma}$. In this parton description, spin liquids appear when the Abrikosov fermions propagate freely. This can be understood in terms of a mean-field decoupling, which yields the free spinon Hamiltonian~\cite{song_doping_2021, kuhlenkamp_tunable_top_2022} 
\begin{equation}
    H_{\text{sp}} = \displaystyle\sum_{\expval{i j} \sigma} J_{i j}^\text{eff} f_{i \sigma}^\dagger f_{j \sigma}.\label{eq:freeSpinons}
\end{equation}
In the chiral spin liquid, the hopping $J_{i j}^\text{eff} =-J\sum_\sigma \langle f_{i\sigma}^\dagger f_{j\sigma} \rangle$ is chosen such that each triangle of the lattice encloses a flux of $\pi/2$, which can be interpreted as a self-generated emergent magnetic field spontaneously breaking time-reversal symmetry. Due to the mean-field decoupling, there are two energy scales: the effective spinon hopping $J^\text{eff} = |J^\text{eff}_{i j}|$, which we keep fixed in our calculation, and the interaction energy between spinons given by Eq. \eqref{eq:HeisenHam}. Both energies scale with $J \sim t^2/U$ and hence decrease with increasing $U$. Still, the relative value of the interaction strength between spinons compared to their effective kinetic energy, $J/J^\text{eff}$, increases as we approach the transition to the $120\degree$ state, implying more fluctuations around the mean-field state. 

We again employ the RPA formalism to describe collective modes, describing quantum fluctuations around the mean-field spinon state through interactions given by Eq.~\eqref{eq:HeisenHam}. In  Refs.~\cite{ghioldi_dynamical_2018, zhang_large-s_2019, zhang_resonating_2020, scheie_proximate_2024} similar approaches were recently applied. Including quantum fluctuations, we find a sharp collective mode below the continuum of two spinon excitations, as shown in panel (1) of Fig.~\ref{fig:3_v1}b). The mode decays into the continuum close to the $\kappa$ point and has a pronounced roton minimum at the $K$ point, indicating short-range $120\degree$ antiferromagnetic correlations. By increasing the effective interaction strength $J$ relative to the spinon hopping $J^\text{eff}$, the roton minimum becomes more pronounced and finally softens at a critical value of $J_c$, as seen in panel (2) of Fig.~\ref{fig:3_v1}b). The closing of the spin gap at the $K$ point indicates an instability towards an ordered phase with ordering wave vector $K$, which is precisely the ordering wave vector for the $120\degree$ phase. Therefore, the excited states of the chiral spin liquid not only carry information about the phase itself but also encode information about competing ordered phases. Using inelastic tunneling spectroscopy, as outlined in Sec.~\ref{sec:inelastic}, we can directly study the mode softening at the transition between the CSL and the $120\degree$ phase. The critical exponent of the gap can hence be directly measured. Experimental observation of the roton mode and its softening would be strong evidence for the spin liquid phase and will reveal correlations relevant for a better understanding of the transition to the ordered state. 

Note that by fixing the spinon hopping $J^\text{eff}_{ij}$, such that there is $\pi/2$ flux per triangle, we assume that the spin liquid remains gapped and neglect modifications of the spinon dispersion as the roton mode closes at the $K$ point. For example, a Dirac spin liquid~\cite{song_unifying_2019} might appear at the critical point or even as an extended critical phase. The Dirac spin liquid is characterized by Dirac cones in the spinon dispersion. In this case, the dynamical spin-structure factor exhibits a continuum of states down to zero frequency at the three distinct $M$ points~\cite{ferrari_dynamical_2019} due to the two Dirac cones, which are separated by lattice momentum $M$. This indicates the fractionalization of spin excitations, while a well-defined roton mode at the $K$ point remains.

One possible challenge for inelastic tunneling spectroscopy is that only the portion of the Brillouin zone, where elastic tunneling is forbidden by energy-momentum conservation, can be measured, see Eq.~\eqref{eq:condIETS}. For the closing of the spin gap, which we predict to occur at the $K$ and $K'$ points, it should always be possible to experimentally tune the chemical potential such that only inelastic tunneling is possible at the points of interest. When the relevant region is closer to the center of the Brillouin zone, this may not be achievable, though. 

In a similar spirit, the QTM can be used to clarify the nature of the transition between the metallic phase and the GWC at fractional filling $\nu = 2/3$ discussed in the previous sections. In Refs.~\cite{li_continuous_2021, ghiotto_quantum_2021}, a continuous metal-insulator transition, driven either by doping or a displacement field, at half-filling was reported. In agreement with Ref.~\cite{watanabe_charge_2005}, we find that the transition from metal to GWC is first order on a mean-field level. The first derivative of the energy and the compressibility, which vanishes for the ordered phase, show a discontinuity at the transition to the metallic phase. By contrast, we observe the charge gap to open continuously, similarly to what was found in an exact diagonalization study~\cite{morales-duran_metal-insulator_2021}, indicating that the transition from metal to insulator is not strongly first order. Further experimental and theoretical studies may help to clarify the situation.

\section{Conclusions and Outlook} \label{sec:conclusion}
This work demonstrates how the QTM can characterize two-dimensional quantum magnets, ranging from ordered (anti-)ferromagnets to quantum spin liquids. In particular, we focus on antiferromagnetic and ferromagnetic generalized Wigner crystals on a triangular moir\'e lattice. There is also the possibility that the nearest-neighbor effective spin coupling vanishes and the state does not magnetically order. In that case, we expect no well-defined spin-wave modes in the inelastic tunneling spectroscopy signal. Note that while the general formulas for the elastic and inelastic tunneling spectroscopy signal, as derived in Appendix~\ref{sec:tunCurrs}, are valid also for finite temperatures, all of our concrete calculations were performed in the zero temperature limit. For concreteness, we focused on a fractional filling of $\nu=2/3$ electrons per site, but GWCs have been observed for a wide range of other fillings~\cite{regan_mott_2020, shimazaki_strongly_2020, xu_correlated_2020, huang_correlated_2021, jin_stripe_2021}. We expect the QTM to distinguish between different types of magnetic order in these other cases as well.

Due to intrinsically strong interactions in atomically thin TMDs, crystalline states can form even in the absence of an induced moir\'e potential~\cite{smolenski_observation_2021, zhou_bilayer_2021}. While such  Wigner crystals have a long history, their spin order, as well as their charge order close to the crystallization transition, are still investigated~\cite{roger_MultipleExchange_1984, Tanatar_GroundState_1989, Zhu_VariationalQuantum_1995, spivak_colloquium_2010, kim_InterstitialInducedFerromagnetism_2022}. Moreover, first-order crystallization transitions are forbidden in this regime due to long-range interactions and replaced by more exotic intermediate phases~\cite{spivak_intermediate_2004, sung_observation_2023, xiang_QuantumMelting_2024}. It will be interesting to generalize our analysis to these systems, as the QTM offers a new opportunity to probe the exotic properties of melting transitions in the two-dimensional electron gases directly. We also remark that systems with weak moir\'e potentials~\cite{shimazaki_strongly_2020} at low fillings may serve as a controlled starting point.

Exotic superconductivity in strongly correlated regimes appears in various two-dimensional heterostructures with and without moir\'e potential~\cite{cao_unconventional_2018, lu_superconductors_2019, zhou_isospin_2022, zerba_realizing_2023}. Often, these superconducting states appear in close proximity to ordered states~\cite{zhou_isospin_2022, Nuckolls_QuantumTextures_2023}. Data from elastic and inelastic scattering via the QTM may help shed light on the collective modes that drive superconductivity in these systems.

Many exotic states, such as Wigner crystals and chiral spin liquids, are stabilized by an external magnetic field~\cite{smolenski_observation_2021, kuhlenkamp_tunable_top_2022}. This makes it interesting to explore the response of the QTM in finite magnetic fields, which has to account for Landau-level formation in the probing graphene layers. A detailed description of the effects of a magnetic field on the QTM is analyzed in Appendix~\ref{sec:mag}~and~\ref{app:4}.

Using the QTM to measure the excitation spectra of the recently discovered fractional quantum anomalous Hall states in twisted $\text{MoTe}_2$ and graphene-based heterostructures~\cite{cai_signatures_2023, zeng_thermodynamic_2023, park_observation_2023, xu_ObservationInteger_2023, lu_FractionalQuantum_2024} offers the exciting possibility of obtaining direct spectroscopic evidence for electron fractionalization.

\textbf{Acknowledgements --} We thank Shahal Ilani, Johannes Knolle, and Josef Willsher for fruitful discussions. We acknowledge support from the Deutsche Forschungsgemeinschaft (DFG, German Research Foundation) under Germany’s Excellence Strategy--EXC--2111--390814868, TRR 360 – 492547816 and DFG grants No. KN1254/1-2, KN1254/2-1, the European Research Council (ERC) under the European Union’s Horizon 2020 research and innovation programme (grant agreement No. 851161), as well as the Munich Quantum Valley, which is supported by the Bavarian state government with funds from the Hightech Agenda Bayern Plus.

\textbf{Data availability --} Data, data analysis, and simulation codes are available upon reasonable request on Zenodo~\cite{zenodo}.

\appendix
\section{Tunneling currents} \label{sec:tunCurrs}
Based on Refs.~\cite{mahan_dc_2000, carrega_tunneling_2020, peri_probing_2024}, we calculate the tunneling current between two layers, described by the independent Hamiltonians $H_T$ and $H_B$ for the top and bottom layer respectively. The two layers are separated by a third intermediate layer, serving as a tunneling barrier. The goal of this section is to derive a unified formalism both for elastic tunneling and inelastic tunneling. In the former case, the top layer is the graphene sheet, the bottom layer is the sample, and the middle layer is an insulating spacer layer that prevents hybridization between the top and bottom layers. In the latter case, the top and bottom layers are graphene sheets, and the middle layer consists of the sample sandwiched between insulating layers. For this setting, we compute the inelastic contributions to the tunneling current. The total Hamiltonian of the system can be written as
\begin{equation}
    H = H_T + H_B + H_{\text{barr}} + H_{\text{tun}}.
\end{equation}
We assume that $H_T$, $H_B$, and $H_{\text{barr}}$ all commute with each other and are only coupled through the tunneling Hamiltonian
\begin{equation}
    H_{\text{tun}} = \displaystyle\sum_{i, j} \displaystyle\sum_{ab, \sigma \sigma'} \left( T^{ab}_{\sigma \sigma'}(\mathbf{x}_i, \mathbf{x}_j) c_{\sigma  a T i}^\dagger c_{\sigma' b B j} e^{i \phi t} + \text{h.c.} \right)
\end{equation}
where $c_{\sigma  a \ell i}^\dagger$ ($c_{\sigma  a \ell i}$) creates (destroys) an electron in layer $\ell \in  \{T, B\}$ on lattice site $\mathbf{x}_i$ with spin $\sigma$ and sublattice $a$. Generically, the (sub-)lattice structure of the top and bottom layers are not assumed to be the same. The bias voltage $\phi$ between the top and bottom layer appears as a time-dependent phase. Assuming that the top and bottom layers are separated by a distance $d$ and that a constant potential $\Phi$ can approximate the tunneling barrier, a WKB approximation yields a tunneling amplitude of
\begin{equation}
    T \sim \exp\left[ -\sqrt{\frac{\Phi-E}{\Phi_0}}\right] \label{eq:tunampl}
\end{equation}
with $\Phi_0 := \frac{8 m d^2}{\hbar^2}$ and $E$ being the energy of the electron. We assume that the electron can couple to spin excitations of the intermediate layer, see Fig.~\ref{fig:2_v1}a), leading to $E= J \mathbf{s}\cdot \boldsymbol{\sigma}$, where $\mathbf{s}$ is the local magnetic moment in the intermediate layer and $J$ is an effective coupling. For small couplings $J/\Phi \ll 1$, the tunneling amplitude Eq.~\eqref{eq:tunampl} can be expanded~\cite{fernandez-rossier_theory_2009, fransson_theory_2010}:
\begin{equation}
    T = \Gamma_0 \mathbb{1} + \Gamma_1 \hat{\mathbf{s}}\cdot \boldsymbol{\sigma} \quad
\end{equation}
with
\begin{equation}
    \frac{\Gamma_1}{\Gamma_0} = \tanh\frac{J |\mathbf{s}|}{2 \Phi} \simeq \frac{J |\mathbf{s}|}{2 \Phi} \ll 1.
\end{equation}
We see two contributions to the tunneling amplitude: a dominating term without spin-flips and a term coupling to spin excitations, suppressed by a relative factor of $\sim J/\Phi$. This motivates the following matrix elements in our tunneling Hamiltonian
\begin{align*}
    T_{\sigma \sigma'}^{ab}(\mathbf{x}_i, \mathbf{x}_j) = \Gamma_0&(\Delta \mathbf{r}_{ij}^{ab}) \delta_{\sigma \sigma'} + \\ &+\Gamma_1(\Delta\mathbf{r}_{ij}^{ab}) \sigma^\mu_{\sigma \sigma'} s^\mu(\mathbf{R}_{ij}^{ab}) \label{eq:tunmatrix} \numberthis
\end{align*}
where $\Delta\mathbf{r}_{ij}^{ab} := (\mathbf{x}_i+\mathbf{r}_a)-(\mathbf{x}_j+\mathbf{r}_b)$ is the relative distance between site $i$ on sublattice $a$ of the top layer and site $j$ on sublattice $b$ of the bottom layer, while $\mathbf{R}_{ij}^{ab}:=(\mathbf{x}_i+\mathbf{r}_a +\mathbf{x}_j+\mathbf{r}_b)/2$ is their midpoint. Note that the first term in Eq.~\eqref{eq:tunmatrix} only depends on the relative distance between the two sites and hence only contributes to elastic tunneling. Inelastic contributions stem from the second term. By taking a Fourier transform 
\begin{align}
    c_{\sigma a \ell i} &= \frac{1}{\sqrt{N}} \displaystyle\sum_{\kk} e^{ i \kk \cdot (\mathbf{x}_i + \mathbf{r}_a)} c_{\sigma a  \ell \kk} \quad \\
    T^{ab}_{\sigma \sigma'}(\Delta \mathbf{r}, \mathbf{R}) &= \frac{1}{N} \displaystyle\sum_{\mathbf{q}, \mathbf{q'}} e^{-i \mathbf{q}\cdot \mathbf{R}} e^{-i \mathbf{q'}\cdot \Delta \mathbf{r}} T^{ab}_{\sigma \sigma'}(\mathbf{q}, \mathbf{q'}) 
\end{align}
we can rewrite the tunneling Hamiltonian as follows

\begin{align*}
    H_{\text{tun}} = &\displaystyle\sum_{ab, \sigma\sigma'} \displaystyle\sum_{\kk \kk'}\displaystyle\sum_{\mathbf{g}_T \mathbf{g}_B} \big(  T^{ab}_{\sigma \sigma'}(\mathbf{q}, \mathbf{q'}) e^{-i (\mathbf{r}_{Ta}\cdot \mathbf{g}_T- \mathbf{r}_{Bb}\cdot \mathbf{g}_B)}\\
    &\times c_{\sigma a T \kk}^\dagger c_{\sigma' b B \kk'} e^{i \phi t} + \text{h.c.} \big) \numberthis
\end{align*}

where $\mathbf{r}_{\ell a}$ is the position of sublattice $a$ within the unit cell of layer $\ell$ and $\mathbf{g}_\ell$ are reciprocal lattice vectors of layer $\ell$. Furthermore, we have $\mathbf{q}= \kk - \kk' + \mathbf{g}_T - \mathbf{g}_B$ and $\mathbf{q}'=(\kk + \kk' + \mathbf{g}_T + \mathbf{g}_B)/2$. The tunneling matrix elements in momentum space can be explicitly written as
\begin{equation}
    T_{\sigma \sigma'}(\mathbf{q}, \mathbf{q'}) = \Gamma_0(\mathbf{q}') \delta_{\mathbf{q}, 0} \delta_{\sigma \sigma'} + \Gamma_1(\mathbf{q}')\sigma^\mu_{\sigma \sigma'} s^\mu_{\mathbf{q}}
\end{equation}
with 
\begin{equation}
    s^\mu_\mathbf{q} = \frac{1}{N} \displaystyle\sum_\mathbf{R} e^{i \mathbf{R}\cdot \mathbf{q}} s^\mu(\mathbf{R}).
\end{equation}
Next, we calculate the tunneling current $I(t) := -e \bexpval{\Dot{N}_T(t)}$ with $N_\ell := \sum_{\kk \sigma a} c^\dagger_{\sigma  \ell a \kk} c_{\sigma \ell a \kk}$. Time-dependent perturbation theory yields
\begin{equation}
    I(t) = -i e \displaystyle\int_{-\infty}^\infty \dd t' \theta(t -t') \bexpval{[H_{\text{tun}}(t'),\; \Dot{N}_T(t)]}
\end{equation}
with  $\Dot{N}_T = i [H_{\text{tun}}, \;  N_T]$. By introducing the bilinear operator 
\begin{align*}
    A(t) := &\displaystyle\sum_{ab, \sigma\sigma'} \displaystyle\sum_{\kk \kk'}\displaystyle\sum_{\mathbf{g}_T \mathbf{g}_B}  T^{ab}_{\sigma \sigma'}(\mathbf{q}, \mathbf{q'}) e^{-i (\mathbf{r}_{Ta}\cdot \mathbf{g}_T- \mathbf{r}_{Bb}\cdot \mathbf{g}_B)} \\
    &\times c_{\sigma a T \kk}^\dagger(t) c_{\sigma' b B \kk'}(t) \numberthis
\end{align*}
one can compactly write
\begin{align*}
    H_{\text{tun}} &= A(t) e^{i \phi t} + A^\dagger(t) e^{-i \phi t}, \\
    \Dot{N}_T(t) &= - i \left[A(t) e^{i\phi t} - A^\dagger(t)e^{-i \phi t}\right], \numberthis
\end{align*}
which yields the single-particle contribution $I_S$ to the tunneling current
\begin{align*}
    I_S(\phi) = e &\displaystyle\int_{-\infty}^\infty \dd t' \theta(t -t') \big[e^{i\phi(t-t')} \expval{[A(t), \; A^\dagger(t')]} \\
    &- e^{-i\phi(t-t')} \expval{[A^\dagger(t), \; A(t')] }  \big] = - 2e \, \Im U_{\text{ret}}(\phi). \numberthis
\end{align*}
We have to calculate the retarded Green's function defined by
\begin{equation}
    U_{\text{ret}}(\omega) = - i \displaystyle\int_{-\infty}^\infty \dd t\; e^{i \omega t} \theta(t) \expval{[A(t), \; A^\dagger(0)]}.
\end{equation}
This can be achieved by first computing the Green's function in the imaginary time formalism
\begin{equation}
    \mathcal{U}(i \omega_n) = - \displaystyle\int_0^{\beta} \dd \tau\; e^{i \omega_n  \tau} \expval{\mathcal{T}A(\tau) A^\dagger(0)}
\end{equation}
and then taking the analytical continuation $i \omega_n \rightarrow \omega + i \eta$ to obtain the retarded Green's function. Here $\mathcal{T}$ is the imaginary time ordering operator. Making use of Wick's theorem and defining the single-particle Green's function as $\mathcal{G}^{ab}_\ell(\kk, i \nu_n) := - \int_0^\beta \dd \tau\; e^{i \nu_n \tau} \bexpval{\mathcal{T} c_{\sigma a \ell \kk}(\tau)c^\dagger_{\sigma b  \ell \kk}(0)}$, we arrive at the following result
\begin{align*}
    \mathcal{U}(i \omega_n) = \frac{1}{\beta^2} &\displaystyle\sum_{\substack{\kk \mathbf{p} \\ \nu_l \Omega_m}}\displaystyle\sum_{\substack{a b \\a' b'}  } \displaystyle\sum_{\alpha \beta} \mathcal{M}^{ab a' b'}_{\alpha\beta}(\kk, \mathbf{p}, i \Omega_m + i\omega_n) \\ &\times\mathcal{G}^{b b'}_B(\kk - \mathbf{p}, i\nu_l -i \Omega_n) \mathcal{G}^{a' a}_T(\kk, i\nu_l) \numberthis \label{eq:UGeneral}
\end{align*}
where $\mathcal{M}^{a b a' b'}_{\alpha \beta}(\kk, \mathbf{p}, i \omega_n)$ is the Fourier transform of
\begin{equation}
    \mathcal{M}^{a ba' b'}_{\alpha \beta}(\kk, \mathbf{p}, \tau) := - \bexpval{\mathcal{T}\, \Tilde{T}^{a b}_{\alpha \beta \kk \mathbf{p}}(\tau)\left(\Tilde{T}^{a' b'}_{\alpha \beta \kk \mathbf{p}}(0)\right)^\dagger } \label{eq:bigM}
\end{equation}
with
\begin{equation}
    \Tilde{T}^{a b}_{\alpha \beta \kk \mathbf{p}}(\tau) = \displaystyle\sum_{\mathbf{g}_T \mathbf{g}_B} T_{\alpha \beta}(\mathbf{q}, \mathbf{q'}; \tau) e^{-i (\mathbf{r}_{Ta}\cdot \mathbf{g}_T- \mathbf{r}_{Bb}\cdot \mathbf{g}_B)} 
\end{equation}
and  $\mathbf{q} = \mathbf{p} + \mathbf{g}_T - \mathbf{g}_B$, $\mathbf{q}'=(2 \kk - \mathbf{p}+ \mathbf{g}_T + \mathbf{g}_B)/2$. 

So far, the theoretical description is general. We will first focus on the elastic tunneling spectroscopy as discussed in Sec.~\ref{sec:elastic} of the main text, which is achieved by setting $\Gamma_1 = 0$. To proceed, we make some simplifying assumptions. Specifically, we assume that the typical in-plane length scale is very small 
compared to the layer-to-layer distance $d$, which implies that $\Gamma_0(\mathbf{q}')$ decays rapidly as a function of $|\mathbf{q'}|$~\cite{bistritzer_transport_2010}. Hence, we can assume it to be constant for the first shell of reciprocal lattice vectors and vanishing otherwise. As in the main text, we focus on the special case where the top layer is a monolayer graphene. For simplicity, we only consider tunneling around a single Dirac point. Thus, for convenience, we redefine momentum to be measured with respect to the $K$ point of the top layer Brillouin zone. Under all of these assumptions, Eq.~\eqref{eq:UGeneral} reduces to
\begin{equation}
    \mathcal{U}^{(0)}(i \omega_n) = \frac{2 |\Gamma_0|^2}{\beta} \displaystyle\sum_{\kk,  \nu_l} \mathcal{G}_B(\kk, i \nu_l + i \omega_n) \mathcal{G}_T(\kk, i \nu_l).
\end{equation}
The Matsubara sum can be performed by making use of the Lehman representation
\begin{equation}
     \mathcal{G}_\ell(\kk, i \omega_n) = \displaystyle\int_{- \infty}^\infty \dd \eps\;  \frac{\mathcal{A}_\ell(\kk, \eps)}{i \omega_n - \eps}
\end{equation}
where $\mathcal{A}_\ell(\kk, \eps) = -\Im \mathcal{G}(\kk, \eps)/\pi$ is the single-particle spectral function of layer $\ell$. After taking the analytical continuation $i \omega_n \rightarrow \omega + i\eta$, we arrive at the final result for the elastic tunneling current
\begin{align*}
    I^{(0)}(\phi) = 4\pi e | \Gamma_0|^2 &\int\dd \eps \displaystyle\sum_\kk \big( f_T(\eps - \phi) - f_B(\eps) \big) \\
    &\times\mathcal{A}_B(\kk, \eps) \mathcal{A}_T(\kk, \eps - \phi) \numberthis
\end{align*}
where $f_\ell(\eps)$ is the Fermi-Dirac distribution for layer $\ell$. 

As a second step, we calculate the inelastic contributions to the tunneling current following Refs.~\cite{carrega_tunneling_2020, peri_probing_2024}, relevant for inelastic tunneling spectroscopy, see Sec.~\ref{sec:inelastic} of the main text. We take both the top and bottom layers to be monolayer graphene, while the insulating barrier is the probe of interest. We again assume that the distance between layers is large compared to the typical length scale of graphene, such that we can restrict ourselves to the first shell of reciprocal lattice vectors and take $\Gamma_1$ to be constant. We only consider contributions from the $K$ valley of graphene for simplicity. The contributions from the $K'$ valley are related by time-reversal symmetry. Under these assumptions, the sum over reciprocal lattice vectors in Eq.~\eqref{eq:bigM} reduces to a sum $n \in \{0, 1, 2\}$ over the three different $K_n$ points as defined in Fig.~\ref{fig:2_v1}a). We introduce the following matrix in sublattice space
\begin{equation}
    T^{(n)}_{ab} = e^{-i (\mathbf{r}_{Ta}\cdot \mathbf{g}^{(n)}_T- \mathbf{r}_{Bb}\cdot \mathbf{g}^{(n)}_B)} 
\end{equation}
which allows us to compactly write the inelastic contribution to Eq.~\eqref{eq:bigM} as
\begin{align*}
    \mathcal{M}^{aba'b'}_{\alpha \beta}(\kk, \mathbf{p}, \tau) = &- |\Gamma_1|^2\displaystyle\sum_{n=0}^2 T^{(n)}_{ab}T^{(n)}_{b'a'} \sigma^{\mu}_{\alpha \beta}\sigma^\nu_{\beta \alpha} \\ 
    &\times\bexpval{\mathcal{T}s^\mu_{\mathbf{p}+\Delta \mathbf{K}_n}(\tau) s^\nu_{-\mathbf{p}-\Delta \mathbf{K}_n}(0)}. \numberthis
\end{align*}
To make further progress, we perform a unitary transformation from sublattice space to band space, where the single-particle Green's functions are diagonal by construction
\begin{align}
    \mathcal{G}^\lambda_\ell(\kk, \tau) \delta_{\lambda \lambda'} \delta_{\kk\kk'} &= \matrixel{\kk, \lambda}{\mathcal{G}^{ab}_\ell(\kk, \tau)}{\kk', \lambda'} \\ 
    T^{(n)}_{\kk\kk', \lambda\lambda'} &= \matrixel{\kk, \lambda}{T^{(n)}}{\kk', \lambda'}
\end{align}
where $\ket{\kk, \lambda}$ is the Bloch-function for graphene in band $\lambda\in\{\pm 1\}$. Inserting all of this back into Eq.~\eqref{eq:UGeneral}, performing the Matsubara sums, and taking the imaginary part after an analytical continuation, we arrive at the following result for the inelastic tunneling current
\begin{align*}
    I^{(2)}(\phi) = &2\pi e |\Gamma_1|^2\displaystyle\sum_{n=0}^2 \int\dd \eps \displaystyle\sum_\kk \big( n_{BE}(\eps + \phi) - n_{BE}(\eps) \big) \\
    &\times \mathcal{S}(\kk +\Delta\mathbf{K}_n, \eps+\phi) \mathcal{A}^{(n)}_{TB}(\kk, \eps) \numberthis \label{eq:inelCurr}
\end{align*}

with $\mathcal{S}(\kk, \eps)$ being the dynamical spin-structure factor defined in Eq.~\eqref{eq:spinspin}, $n_{BE}(\eps)$ the Bose-Einstein distribution and 
\begin{align*}
    \mathcal{A}_{TB}^{(n)}(\kk, \eps) = &\displaystyle\sum_{\mathbf{q}, \lambda\lambda'} \int \dd \omega \big( f_T(\omega) - f_B(\omega-\eps))| T^{(n)}_{\mathbf{q}, \mathbf{q}+\kk, \lambda\lambda'}|^2 \\ &\times  \mathcal{A}_T^\lambda(\mathbf{q}, \eps) \mathcal{A}_B^{\lambda'}(\mathbf{q}+\kk, \omega-\eps). \numberthis \label{eq:ATB}
\end{align*}
Here $\mathcal{A}_\ell^\lambda(\mathbf{q}, \eps) = - \Im \mathcal{G}_\ell^\lambda(\mathbf{q}, \eps)/\pi$ is again the single-particle spectral function of graphene for layer $\ell$ and band $\lambda$. To further simplify the convolution of Eq.~\eqref{eq:inelCurr}, we follow the same steps as outlined in Ref.~\cite{peri_probing_2024}. Taking the zero-temperature limit, both the difference between Bose-Einstein and Fermi-Dirac distributions, in Eq.~\eqref{eq:inelCurr} and~\eqref{eq:ATB}, reduce to theta-functions limiting the integration boundaries. Next, we make use of the fact that the Fermi-velocity $v_F$ in graphene is very large compared to typical moir\'e scales $\hbar v_F a_\text{gr} \gg \mu \theta$, where $a_\text{gr}$ is the lattice constant of graphene and $\theta$ is the twist angle between the top and bottom layer. Under this assumption, we can approximate $\mathcal{S}(\kk +\Delta\mathbf{K}_n, \eps) \approx \mathcal{S}(\Delta\mathbf{K}_n, \eps)$. Furthermore, we assume that the chemical potentials of the top and bottom layers are large compared to the bias voltage between them $\mu_T = \mu_B \equiv \mu \gg \phi$. This allows us to assume that the density of states $\rho(\eps) = \sum_{\kk \lambda} \mathcal{A}^\lambda(\kk, \eps)/N \sim |\mu + \eps| \approx |\mu|$ in graphene is nearly constant for $\eps < \phi$. Finally, we replace the matrix elements $| T^{(n)}_{\mathbf{q}, \mathbf{q}+\kk, \lambda\lambda'}|^2$ by their average, which is unity. Under these assumptions, all momentum sums and one of the energy integrals can be carried out explicitly, leading to Eq.~\eqref{eq:inelSIM}. 

In our initial motivation for the tunneling matrix Eq.~\eqref{eq:tunmatrix}, we assumed that the inelastic contributions come from interactions between the tunneling electron and low-energy spin excitations in the intermediate layer. If we probe at higher energies instead, we can couple to density excitations. Concretely, this would correspond to a tunneling matrix $T_{\sigma \sigma'} \sim \delta_{\sigma \sigma'} \rho(\mathbf{R})$, where $\rho(\mathbf{R})$ is the local density in the intermediate layer. In that case, one measures the density response of Eq.~\eqref{eq:denden} instead. 

There are also mixed contributions to the current, proportional to $\Gamma_0 \Gamma_1$, but when coupling to spin excitations they vanish since $\mathcal{M}^{aba'b'}_{\alpha \beta} \sim \sigma^\mu_{\alpha \beta} $ and consequently the spin sum in Eq.~\eqref{eq:UGeneral} is zero because $\Tr\sigma^\mu = 0$.

\section{Mean-field theory} \label{sec:MFT}
The matrix $h_\kk$ in spinor space in Eq.~\eqref{eq:bilHam} is given by
\begin{align*}
    [h_\kk^{(t)}]_{ab, \sigma\sigma'} &= -t \gamma^{ab}_\kk \delta_{\sigma \sigma'} \numberthis\\
    [h_\kk^{(u)}]_{ab, \sigma\sigma'} &= U \delta_{a b} \mfp_{\sigma \sigma'}^{a b}( \delta_{\sigma \sigma'} - \sigma_{\sigma \sigma'}^x) \numberthis \\
    [h_\kk^{(v)}]_{ab, \sigma\sigma'} &= V\big[3 (n - n_a)\delta_{a b} \delta_{\sigma \sigma'} -\gamma^{a b}_\kk \mfp^{b a}_{\sigma' \sigma} \big]\numberthis\\
    [h_\kk^{(x)}]_{ab, \sigma\sigma'} &= X\big[ 3\big(\mfp_{\sigma'\sigma}^{ab} - \displaystyle\sum_c \mfp_{\sigma' \sigma}^{c c}\big)\delta_{ab}
    + \\ 
    &\quad\quad\quad\quad\quad\quad +\gamma_\kk^{ab} \delta_{\sigma \sigma'} \displaystyle\sum_{\alpha} \mfp_{\alpha\alpha}^{ba}\big] \numberthis
\end{align*}
with $n_a = \mfp^{a a}_{\uparrow \uparrow} + \mfp^{a a}_{\downarrow \downarrow}$ and $n = \sum_a n_a$. For the nearest-neighbor geometric factor $\gamma_\kk^{ab}$ one finds
\begin{align}
    \gamma_\kk^{AB} &= 1 + e^{i \mathbf{b}_1\cdot \kk} + e^{i (\mathbf{b}_1 + \mathbf{b}_2)\cdot \kk} \\
    \gamma_\kk^{AC} &= 1 + e^{i \mathbf{b}_2\cdot \kk} + e^{i (\mathbf{b}_1 + \mathbf{b}_2)\cdot \kk} \\
    \gamma_\kk^{BC} &= 1 + e^{-i \mathbf{b}_1\cdot \kk} + e^{i \mathbf{b}_2\cdot \kk}
\end{align}
and $\gamma_\kk^{aa}=0 \;\; \forall\; a \in \{A, B, C\}$, $\gamma_\kk^{ba} = (\gamma_\kk^{ab})^*$. The vectors $\mathbf{b}_{1/2}$ span the unit cell and are defined in Fig.~\ref{fig:1_v2}b). The following constant energy shift is obtained
\begin{align*}
    E_0 = &- U N \displaystyle\sum_a \big( \mfp_{\uparrow \uparrow}^{a a} \mfp_{\downarrow \downarrow}^{a a} - \mfp_{\uparrow \downarrow}^{a a} \mfp_{\downarrow \uparrow}^{a a}\big) + \\ &+ \frac{V N}{2} \displaystyle\sum_{a b}\displaystyle\sum_{\sigma \sigma'} \gamma_{\kk=0}^{ab}\big(|\mfp_{\sigma \sigma'}^{a b}|^2 - \mfp_{\sigma \sigma}^{a a} \mfp_{\sigma' \sigma'}^{b b} \big) + \\ &+ \frac{X N}{2} \displaystyle\sum_{a b, \sigma} \gamma_{\kk=0}^{ab}\big(\mfp_{\sigma \sigma}^{a b}\mfp_{\Bar{\sigma}\Bar{\sigma}}^{b a} - \mfp_{\Bar{\sigma} \sigma}^{a a} \mfp_{\sigma \Bar{\sigma}}^{b b}\big). \numberthis
\end{align*}

The mean-field result for the retarded single-particle Green's function of Eq.~\eqref{eq:GF} is given by
\begin{equation}
    \mathcal{G}^{(0)}(\kk, \omega) = \displaystyle\sum_{\mu=1}^6 \frac{Z^\mu_\kk}{\omega - \eps_\kk^\mu + i \eta},
\end{equation}
where $\eps_\kk^\mu$ are the eigenvalues of the mean-field matrix $h_\kk$ and $Z^\mu_\kk$ is the following function of the eigenvector projections $\alpha_{a \sigma \kk}^\mu$
\begin{equation}
    Z_\kk^\mu = \displaystyle\sum_{\sigma, a b} e^{- i \kk \cdot (\mathbf{r}_a - \mathbf{r}_b)} \alpha_{a \sigma \kk}^\mu (\alpha_{b \sigma \kk}^\mu)^*
    \end{equation}
with $\mathbf{r}_a$ describing the position of sublattice $a$ within the unit cell.

\section{Effective hopping in MF background} \label{sec:eff_hop}
In the AFM, the dispersion of both particle bands and the hole band is described as a free particle, with hopping restricted to the initial sublattice it has been placed on since the other two sublattices are inaccessible either due to energy conservation or the Pauli principle.  

The effective hopping parameters can be computed using perturbation theory for $U \gg V$ and $X=0$. For particles hopping on a triangular lattice spanned by the two vectors $\mathbf{b}_{1/2}$, they are given by
\begin{align}
    t^{(\text{eff})}_1 &= \frac{3t^2}{6 V} + \frac{3 t^2}{2(3V-U)} + \frac{18 t^3}{3V (3V-U)} + \cdots \\
    t^{(\text{eff})}_2 &= -\frac{3t^2}{2 U} - \frac{3 t^2}{2(U-3V)} + \frac{18 t^3}{U (U-3V)} + \cdots \\
    t^{(\text{eff})}_0 &= \frac{3 t^2}{6 V} + \frac{3 t^2}{2 U} + \frac{6 t^3}{U V} + \cdots 
\end{align}
where $t^{(\text{eff})}_1$ and $t^{(\text{eff})}_2$ correspond to the upper and lower particle band respectively and $t^{(\text{eff})}_0$ describes the hole band.
For the ferromagnet, the two topmost particle bands and the two hole bands correspond to leading order to free hopping on a hexagonal lattice with hopping parameter $t$. The remaining two particle bands can be understood as free hopping on a triangular lattice and effective hopping parameters
\begin{align}
    t^{(\text{eff})}_1 &= -\frac{2t^2}{2 V- X} + \frac{4 t^3}{(2V-X)^2} +  \cdots \\
    t^{(\text{eff})}_2 &= \frac{6t^2}{2 U-6V-3X} +\frac{36 t^3}{(2 U-6V-3X)^2} + \cdots 
\end{align}
assuming $U \gg V, X$.

\section{Random Phase Approximation} \label{sec:RPA}
To incorporate quantum fluctuations on top of mean-field states, we use the random phase approximation (RPA) to calculate the spin- and density response functions. We consider a generic interaction Hamiltonian for fermions
\begin{equation}
    H_{\text{int}} = \frac{1}{N}\displaystyle\sum_{\kk, \kk', \qb} \displaystyle\sum_{\alpha \beta \gamma \delta} V_{\alpha\beta\gamma\delta}(\qb) c^\dagger_{\alpha \kk+\qb} c^\dagger_{\beta \kk'-\qb} c_{\gamma \kk'} c_{\delta \kk} \label{eq:RPA_int}
\end{equation}
where Greek indices can be both spin and sublattice, and $N$ is the number of unit cells. We calculate the generic response function
\begin{align*}
    \chi_{\alpha \beta \gamma \delta}(\qb, i \omega_n) = &\frac{1}{2} \displaystyle\int_0^\beta \dd \tau\; e^{i \omega_n \tau}\\
    &\times \displaystyle\sum_{\kk, \kk'} \big\langle{\mathcal{T}c^\dagger_{\alpha \kk}(\tau) c_{\beta \kk+\qb}(\tau) c^\dagger_{\gamma \kk'}(0) c_{\delta \kk'-\qb}(0)\big\rangle}, \numberthis
\end{align*}
which can be expressed in terms of the single-particle Green's function $\mathcal{G}_{\alpha \beta}(\kk, \tau) = - \bexpval{\mathcal{T}c_{\alpha \kk}(\tau)c^\dagger_{\beta \kk}(0)}$ as
\begin{equation}
    \chi_{\alpha \beta \gamma \delta}(\qb, i \omega_n) = - \frac{1}{2 \beta} \displaystyle\sum_{\kk, i\nu_m} \mathcal{G}_{\delta \alpha}(\kk, i \nu_m) \mathcal{G}_{\beta \gamma}(\kk + \qb, i \omega_n + i \nu_m).
\end{equation}
In the RPA formalism, the response function fulfills the Dyson equation
\begin{align*}
    \chi_{\alpha\beta\gamma\delta}(\qb, i \omega_n) = &\displaystyle\sum_{\alpha'\beta'\gamma'\delta'}\chi_{\alpha\beta\alpha'\beta'}(\qb, i \omega_n) V_{\alpha' \beta' \gamma' \delta'}(\qb) \\
    &\times \chi^{(0)}_{\gamma'\delta'\gamma\delta}(\qb, i \omega_n) + \chi_{\alpha\beta\gamma\delta}^{(0)}(\qb, i \omega_n), \numberthis
\end{align*}

which has the following solution in matrix form
\begin{equation}
    \cchi_q = (\mathbb{1} - \cchi^{(0)}_q \mathbf{V}_\qb)^{-1} \cchi^{(0)}_q, \label{eq:RPAsolution}
\end{equation}
where we have introduced a combined energy-momentum label $q=(\qb, i \omega_n)$. The bare response function $\cchi^{(0)}_q$ is computed on a mean-field level:
\begin{align*}
    \chi_{\alpha \beta \gamma \delta}^{(0)}(\qb, \omega) = &\frac{1}{2 N} \displaystyle\sum_\kk \displaystyle\sum_{\mu \nu} \frac{n^\mu_{\kk+\qb}-n^\nu_\kk}{\omega + \eps^\mu_{\kk+\qb}-\eps^\nu_\kk + i\eta} \\ &\times(\alpha^\mu_{\alpha, \kk+\qb})^* \alpha^\nu_{\beta \kk}\alpha^\mu_{\delta, \kk+\qb}(\alpha^\nu_{\gamma \kk})^* \numberthis\label{eq:bareResponse}
\end{align*}
with $\alpha_{\beta \kk}^\mu$ being the projection of the eigenstate with eigenvalue $\eps_\kk^\mu$. Finally, the dynamical spin-structure factor is given by
\begin{align*}
    \mathcal{S}^{i j}(\kk, \omega) = &-\frac{1}{\pi} \Im  \displaystyle\sum_{ab} \displaystyle\sum_{\alpha \beta \gamma \delta} e^{- i \kk \cdot (\mathbf{r}_a - \mathbf{r}_b)} \\ &\times \chi_{aa bb}^{\alpha\beta\gamma\delta}(\kk, \omega) \sigma_{\alpha \beta}^i \sigma_{\gamma\delta}^j, \numberthis
\end{align*}
where we have explicitly introduced sublattice indices $a,b$. For the spin response in the ordered $120\degree$ phase and the antiferromagnetic generalized Wigner crystal, we use an on-site interaction for the RPA calculation. For the chiral spin liquid phase, we work with an antiferromagnetic Heisenberg-Hamiltonian, which can be expressed in terms of Abrikosov fermions as Eq.~\eqref{eq:HeisenHam}. After performing a mean-field decoupling, one gets the free spinon Hamiltonian of Eq.~\eqref{eq:freeSpinons}. We assume a Kalmeyer-Laughlin type of chiral spin liquid~\cite{kalmeyer_equivalence_1987}, which fixes the hoppings $J^\text{eff}_{ij}$ up to a gauge transformation. Using the free spinon Hamiltonian of Eq.~\eqref{eq:freeSpinons}, we compute the bare response and then apply the above-described RPA formalism, expressing the interaction Hamiltonian Eq.~\eqref{eq:HeisenHam} in momentum space as
\begin{equation}
    H_{\text{int}} = \frac{1}{N} \displaystyle\sum_{\substack{\kk, \kk', \qb \\ \alpha \beta}}\displaystyle V(\qb) f^\dagger_{\alpha \kk+\qb} f^\dagger_{\beta \kk'-\qb} f_{\alpha \kk'} f_{\beta \kk} 
\end{equation}
with 
\begin{equation}
    V(\qb) = \frac{J}{4} \displaystyle\sum_{n=0}^5 e^{-i \qb \cdot \ddelta_n}
\end{equation}
where $\ddelta_n = \mathrm{R}\left(\frac{n \pi}{3}\right) \ddelta_0$, $\ddelta_0 = (1,0)^T$ are the six nearest neighbors in the triangular lattice  and $\mathrm{R}(\theta)$ is a rotation matrix.

\section{Spin-wave dispersion from Holstein-Primakoff transformation}  \label{sec:HPdisp}
Consider the $J_1 - J_2$ model on a hexagonal lattice
\begin{equation}
    H = J_1 \displaystyle\sum_{\expval{i j}} \mathbf{S}_i \cdot \mathbf{S}_j + J_2 \displaystyle\sum_{\expval{\expval{i j}}}\mathbf{S}_i \cdot \mathbf{S}_j
\end{equation}
with nearest-neighbor spin coupling $J_1$ and next-nearest-neighbor spin coupling $J_2$. We assume that $| J_1| \gg |J_2|$ is the dominant energy scale. Explicitly in terms of sublattices
\begin{align*}
    H = &J_1 \displaystyle\sum_\mathbf{i} \displaystyle\sum_{\boldsymbol{\delta}=1}^3 \mathbf{S}_{A, \mathbf{i}} \cdot  \mathbf{S}_{B, \mathbf{i} + \boldsymbol{\delta}} + \\ &+J_2 \displaystyle\sum_{\mathbf{i}} \displaystyle\sum_{\boldsymbol{\delta}'=1}^3 (\mathbf{S}_{A, \mathbf{i}} \cdot  \mathbf{S}_{A, \mathbf{i} + \boldsymbol{\delta}'} + \mathbf{S}_{B, \mathbf{i}} \cdot  \mathbf{S}_{B, \mathbf{i} + \boldsymbol{\delta}'}) \numberthis
\end{align*}
where $\mathbf{S}_{a, \mathbf{i}}$ is the spin operator on sublattice $a \in \{A, B\}$ and site $\mathbf{i} \equiv  \mathbf{x}_i$. The nearest-neighbor vectors $\ddelta$ and $\ddelta'$ are chosen as follows
\begin{align}
    &\ddelta_1 = 0, \quad \ddelta_2 = -\mathbf{b}_1, \quad \ddelta_3 = -\mathbf{b}_1 - \mathbf{b}_2 \\
    &\ddelta'_1 = \mathbf{b}_1, \quad \ddelta'_2 = \mathbf{b}_2, \quad \ddelta'_3 = \mathbf{b}_1 + \mathbf{b}_2 
\end{align}
with $\mathbf{b}_{1/2}$ defined in Fig.~\ref{fig:1_v2}b). 

We consider the ferromagnetic case $J_1 < 0$. Here the Holstein-Primakoff transformation for sublattice $A$ and $B$ is equivalent:
\begin{align}
    S^x_{a, \mathbf{i}} &= S - a_\ii^\dagger a_\ii \\
    S^+_{a, \ii} &= \sqrt{2 S} \sqrt{1 - \frac{a_\ii^\dagger a_\ii}{2 S}} a_\ii \\
    S^-_{a, \ii} &= \sqrt{2 S} a_\ii^\dagger \sqrt{1 - \frac{a_\ii^\dagger a_\ii}{2 S}} 
\end{align}
with $S^\pm = S^y \pm i S^z$. The bosonic creation and annihilation operators describe spin excitations around the ordered ground state, defined by $a_\ii \ket{0} = 0$. We expand in the total spin $S \gg 1$, even though we are eventually interested in the $S=1/2$ case. The resulting quadratic spin-wave Hamiltonian is given in momentum space by
\begin{align}
    H &= -3 J_1 S \displaystyle\sum_\kk (a_\kk^\dagger,  b_\kk^\dagger) H(\kk) \vc{a_\kk}{b_\kk}\\
    H(\kk) &= 
 \begin{pmatrix}
  1-\frac{2 J_2}{J_1}(1-\xi_\kk) & -\gamma_\kk \\
  -\gamma_{\kk}^* & 1-\frac{2 J_2}{J_1}(1-\xi_\kk) 
 \end{pmatrix} \label{eq:Ham_k_FM}
\end{align}
with $\xi_\kk = 1/3 \sum_{\ddelta'} \cos(\kk \cdot \ddelta')$ and $\gamma_\kk = 1/3 \sum_{\ddelta} e^{- i \kk \cdot \ddelta}$. Diagonalizing $H(\kk)$ yields the spin-wave dispersion
\begin{equation}
    \eps_\kk^\pm = 3 J_1 S \big[ 1 - \frac{2 J_2}{J_1}(1-\xi_\kk) \pm |\gamma_\kk| \big].
\end{equation}
The lower branch is gapless at $\kk=0$ and the $K$ and $K'$ points of the Brillouin zone, while the upper branch is gapped in the long-wavelength limit. The gap between the two branches closes at finite energy at the corners $\kappa$ of the reduced Brillouin zone, forming bosonic Dirac cones~\cite{pershoguba_dirac_2018}. For $J_2=0$, the two bands have the same bandwidth. For antiferromagnetic $J_2 < 0$, the bandwidth of the lower branch increases, while the bandwidth of the upper branch decreases. For ferromagnetic $J_2 >0$, the opposite is the case. Consequently, the main effect of a finite next-nearest-neighbor spin coupling is to break the symmetry between the two branches by renormalizing the energies differently.  
In the basis that diagonalizes the Hamiltonian of Eq.~\eqref{eq:Ham_k_FM}, the magnon operators can be written as
\begin{equation}
    a_\kk = \displaystyle\sum_{\lambda\in\{\pm\}} u^a_{\lambda \kk} f_{\lambda \kk}.
\end{equation}
In terms of the coefficients $u^a_{\lambda \kk}$, the spin-spin response 
\begin{equation}
    \chi^{\mu \nu}_{a b}(\kk,  \omega) = - \frac{i}{N} \displaystyle\int_0^{\infty} \dd t \; e^{i \omega t} \expval{[S^\mu_{a, \kk} (t), S^{\nu}_{b, -\kk}(0)]}
\end{equation}
is given by
\begin{align*}
    \chi_{a b}^{xx}(\kk, \omega) &= 0, \numberthis \\
    \chi_{a b}^{yy}(\kk, \omega) &= \chi_{ab}^{zz}(\kk, \omega) \\
    &= \frac{S}{2}\displaystyle\sum_{\lambda\in\{\pm\}} \left[ \frac{u_{\lambda, -\kk}^a (u_{\lambda, -\kk}^b)^*}{\omega - \eps_\kk^\lambda + i \eta} - \frac{(u_{\lambda, \kk}^a)^* u_{\lambda, \kk}^b}{\omega + \eps_\kk^\lambda + i \eta}\right] \numberthis \\
    \chi_{a b}^{y z}(\kk, \omega) &= - \chi_{a b}^{y z}(\kk, \omega) \\
    &= - i \frac{S}{2 } \displaystyle\sum_{\lambda\in\{\pm\}} \left[ \frac{u_{\lambda, -\kk}^a (u_{\lambda, -\kk}^b)^*}{\omega - \eps_\kk^\lambda + i \eta} + \frac{(u_{\lambda, \kk}^a)^* u_{\lambda, \kk}^b}{\omega + \eps_\kk^\lambda + i \eta}\right]. \numberthis
\end{align*}

The total spin-structure factor is
\begin{equation}
    \mathcal{S}(\kk, \omega) = -\frac{1}{\pi} \Im \displaystyle\sum_\mu \displaystyle\sum_{ab} e^{- i \kk \cdot (\mathbf{r}_a - \mathbf{r}_b)} \chi_{ab}^{\mu \mu}(\kk, \omega). 
\end{equation}

\section{Tunneling in an external magnetic field} \label{sec:mag}

\begin{figure}
\begin{center}
\includegraphics[width=\linewidth]{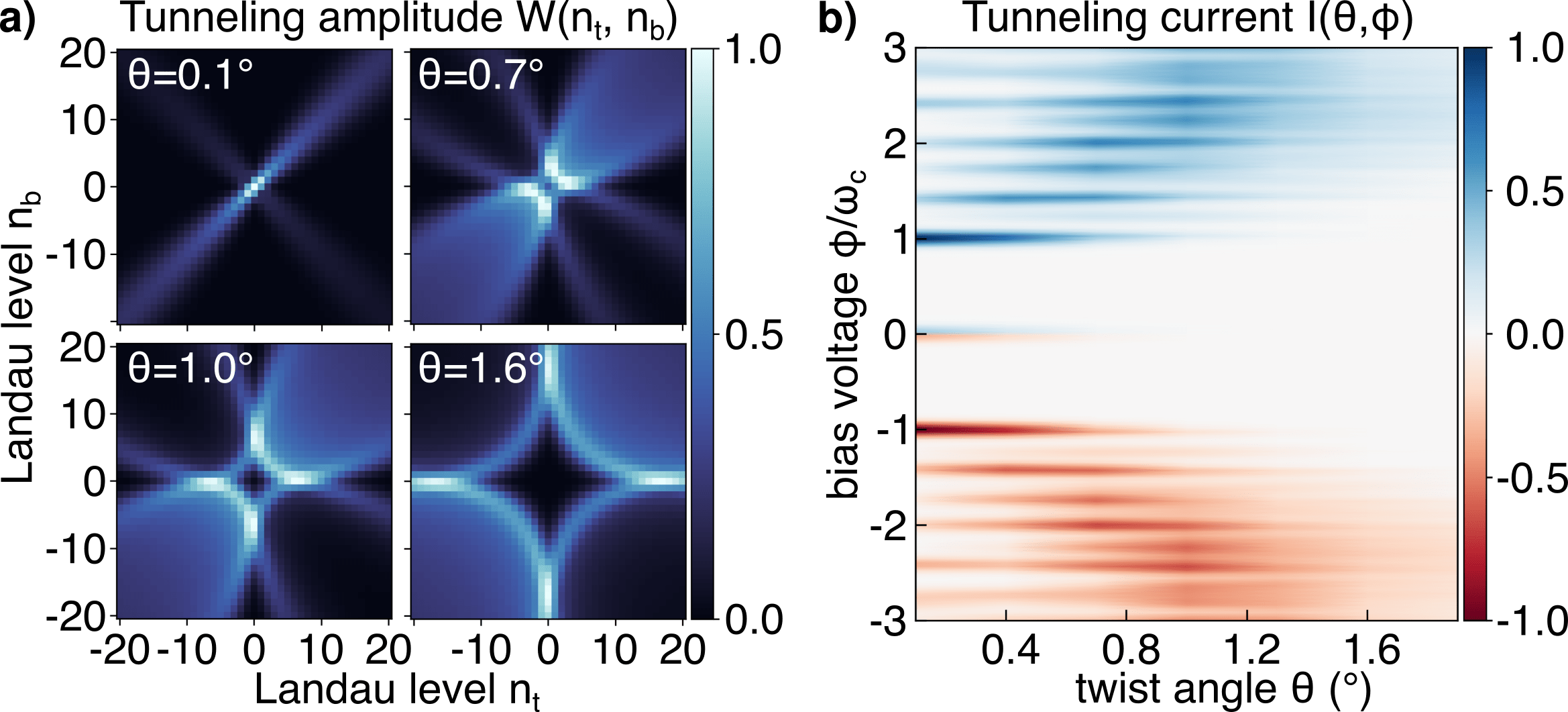}
\caption{\textbf{Elastic tunneling in a magnetic field.} An external magnetic field (here $B=4$ T) leads to the formation of relativistic Landau levels in the top and bottom layer graphene. \textbf{a)} Normalized tunneling amplitudes between Landau level $n_\tfr$ in the top layer and $n_\bfr$ in the bottom layer for different twist angles. \textbf{b)} We consider elastic tunneling between two sheets of graphene separated by an insulating barrier. A bias voltage is applied between the graphene layers, and the system is placed in a perpendicular magnetic field. We show the elastic tunneling current as a function of the bias voltage $\phi$ and the twist angle $\theta$. Both graphene layers are at charge neutrality.}
\label{fig:magfield_el}
\end{center}
\end{figure}

The chiral spin liquid phase encoded in the layer pseudo-spin can be stabilized by an external magnetic field~\cite{kuhlenkamp_tunable_top_2022}. Therefore, we explore the response of the QTM in finite magnetic fields, which has to account for Landau-level formation in the probing graphene layers. We consider a similar setup to the one discussed in Sec.~\ref{sec:inelastic}: The tunneling current between two graphene sheets is measured as a function of the applied bias voltage and the relative twist angle, with the probe of interest sandwiched between the two graphene layers. Additionally, we now apply an external perpendicular magnetic field. This introduces a new length scale: the magnetic length $\ell_B = \sqrt{\hbar/eB}$, which limits momentum resolution by $\delta k \sim 1/\ell_B \sim \sqrt{B}$. Consequently, we expect that for sufficiently large magnetic fields, when the magnetic length is of the same order as the typical length scale of our probe, the quantum twisting microscope completely loses its momentum resolution. Intuitively, this can be understood as a consequence of the flattening of the Dirac cones in graphene due to the formation of relativistic Landau levels. Ideally, there will be an intermediate regime where the momentum space resolution is still sufficiently good to uniquely characterize the dynamical response functions of the probe while already stabilizing the phase of interest. 

The low energy description of the $K = K^+$ ($K'=K^-$) valley of a monolayer graphene in an external gauge potential $\mathbf{A}(\mathbf{r})$ is given by~\cite{Goerbig_ElectronicProperties_2011, greenaway_ResonantTunnelling_2015}
\begin{equation}
    H_{K^\pm} = v_F ( \pm \Pi_x \sigma^x + \Pi_y \sigma^y) \quad \text{with} \quad \boldsymbol{\Pi} = \mathbf{p} + e \mathbf{A} \label{eq:maggraph}
\end{equation}
Following Ref.~\cite{greenaway_ResonantTunnelling_2015}, we introduce the relative twist angle $\theta$ between the two layers by a constant pseudo vector potential, which rotates the momenta. We adopt the following gauge
\begin{equation}
    \mathbf{A}_\ell(\mathbf{r}) = \frac{1}{e} \big( \ell \Delta K^\pm_x, -e B x + \ell \Delta K^\pm_y, 0 \big)^T
\end{equation}
with the layer index $\ell \in \{ \bfr \hat{=} 0, \tfr \hat{=} 1\}$. The eigenvalues of Eq.~\eqref{eq:maggraph} are given by
\begin{equation}
    \eps_n = \text{sgn}(n) \omega_c \sqrt{n}
\end{equation}
with cyclotron frequency $\omega_c=\sqrt{2eB}v_F$ and the relativistic Landau-level index $n \in \mathbb{Z}$. The corresponding eigenstates are~\cite{greenaway_ResonantTunnelling_2015}

\begin{align*}
    \psi^+_{n, k_y}(\mathbf{r}) &= \frac{C_n}{\sqrt{L}} e^{i k_y y} \vc{\phi^{(\ell)}_{|n|}(x)}{-\text{sgn}(n) i \phi^{(\ell)}_{|n|-1}(x)}, \\
    \psi^-_{n, k_y}(\mathbf{r}) &= \frac{C_n}{\sqrt{L}} e^{i k_y y} \vc{\text{sgn}(n) i \phi^{(\ell)}_{|n|-1}(x)}{\phi^{(\ell)}_{|n|}(x)} \numberthis \label{eq:LLeigenstates}
\end{align*}
with $C_n = \sqrt{(1+\delta_{n, 0})/2}$ and 
\begin{align*}
    \phi^{(\ell)}_n(x) = &\frac{1}{\sqrt{2^n n! \sqrt{\pi} \ell_B}} H_n\big(\frac{x-X_\ell}{\ell_B}\big)\\
    &\times \exp\left[-\frac{(x-X_\ell)^2}{2 \ell_B^2} - i \ell \Delta K ^\pm_x (x-X_\ell) \right],  \numberthis\label{eq:phistates}
\end{align*}
where $H_n(x)$ are Hermite polynomials and $X_\ell = \ell_B^2 (k_y + \ell \Delta K^\pm_y)$ is the $x$-component of the orbit center for the semiclassical cyclotron orbits. The quantum number $k_y$ is indexing the degeneracy of each Landau level. 
The derivation of the tunneling current is analogous to the one presented in Sec.~\ref{sec:elastic} and Sec.~\ref{sec:inelastic}. We again assume that all layers are separated by some insulating barrier and only coupled through a tunneling Hamiltonian 
\begin{equation}
    H_\text{tun} = \displaystyle\sum_{\mathbf{r}_\tfr, \mathbf{r}_\bfr}\displaystyle\sum_{a b}\big( T_{a b}(\mathbf{r}_\tfr, \mathbf{r}_\bfr) c^\dagger_{\tfr a}(\mathbf{r}_\tfr) c_{\bfr a}(\mathbf{r}_\bfr)e^{i \phi t} + \text{h.c.} \big).
\end{equation}
We take tunneling matrix elements of the form
\begin{equation}
    T_{ab}(\mathbf{r}_\tfr, \mathbf{r}_\bfr) = \hat{T}_{ab} \delta(\mathbf{r}_\tfr - \mathbf{r}_\bfr) \big[ \Gamma_0 + \Gamma_1 \hat{O}(\mathbf{r}_\tfr) \big] 
\end{equation}
where $\hat{O}(\mathbf{r})$ is some operator that the electrons couple to when tunneling through the intermediate layer. Concretely, this operator might be the local magnetic moment or the local density, as discussed in Appendix~\ref{sec:tunCurrs}. For simplicity, we take the tunneling amplitude between all sublattices to be equal, i.e., $\hat{T}_{ab} \sim \delta_{ab} + \sigma^x_{a b}$. 
First, we focus on the elastic contribution to the tunneling current, setting $\Gamma_1 = 0$. In the following, we restrict ourselves to the $K$ valley. Analogous expressions can be derived for the $K'$ valley. In the basis of the Landau level eigenfunction given by Eq.~\eqref{eq:LLeigenstates}, the elastic tunneling Hamiltonian can be expressed as
\begin{equation}
    H^{(0)}_\text{tun} = \displaystyle\sum_{n_\tfr, n_\bfr}\displaystyle\sum_{k_\tfr, k_\bfr}\big( T_{n_\tfr n_\bfr k_\tfr k_\bfr} c^\dagger_{\tfr n_\tfr k_\tfr} c_{\bfr n_\bfr k_\bfr} e ^{i \phi t} + \text{h.c.}\big)
\end{equation}
with 
\begin{equation}
    T_{n_\tfr n_\bfr k_\tfr k_\bfr} = \Gamma_0 \displaystyle\sum_{\mathbf{r}} \psi^*_{n_\tfr, k_\tfr}(\mathbf{r}) \hat{T} \psi_{n_\bfr, k_\bfr}(\mathbf{r}).
\end{equation}
A similar calculation to the one presented in Sec.~\ref{sec:elastic}, yields an elastic tunneling current of the form
\begin{align*}
    I^{(0)}(\phi) = &- 2\pi e \displaystyle\sum_{n_\tfr, n_\bfr}\displaystyle\sum_{k_\tfr, k_\bfr} |T_{n_\tfr n_\bfr k_\tfr k_\bfr}|^2 \\
    &\times \int \dd \eps \; \big[ f_\tfr(\eps) - f_\bfr(\eps+\phi) \big] \\
    &\times \mathcal{A}_\tfr(\eps, n_\tfr) \mathcal{A}_\bfr(\eps+\phi, n_\bfr). \numberthis
\end{align*}
where $\mathcal{A}_\ell(\eps, n)$ is the spectral function of layer $\ell$ and Landau level $n$. In the disorder-free limit we can take $\mathcal{A}_\ell(\eps, n) \sim \delta(\eps-\eps_n)$, which results in a current
\begin{align*}
    I^{(0)}(\phi) = &- 2\pi e \displaystyle\sum_{n_\tfr, n_\bfr} W(n_\tfr, n_\bfr)\big[ f_\tfr(\eps_{n_\tfr}) - f_\bfr(\eps_{n_\tfr}+\phi) \big] \\
    &\times \delta(\eps_{n_\tfr}-\eps_{n_\bfr}+\phi). \numberthis \label{eq:magcurr_elast}
\end{align*}

The dependence on the twist angle is implicitly contained in the tunneling amplitudes
\begin{equation}
    W(n_\tfr, n_\bfr) = \frac{L^4}{(2\pi)^2} \int \dd k_\tfr \dd k_\bfr \; |T_{n_\tfr n_\bfr k_\tfr k_\bfr}|^2 \label{eq:tunampl_el}
\end{equation}
which describe the probability for an electron in the top layer in Landau level $n_\tfr$ to tunnel to Landau level $n_\bfr$ in the bottom layer. Explicit expressions for $W(n_\tfr, n_\bfr)$ are given in Appendix~\ref{app:4}. 
In Fig.~\ref{fig:magfield_el}a), we plot $W(n_\tfr, n_\bfr)$ for four different twist angles at a fixed magnetic field. 
For very small twist angles, the tunneling amplitude is peaked at transitions between the same Landau levels $|n_\tfr| = |n_\bfr|$. As we move to higher twist angles, the maxima of $W(n_\tfr, n_\bfr)$ are shifted towards transitions between increasingly separated Landau levels. This observation can be explained by a simple semiclassical picture~\cite{greenaway_ResonantTunnelling_2015}: the massless Dirac fermions of graphene perform cyclotron orbits in the presence of a magnetic field. The circular motion in real space induces orbital motion also in momentum space, with an orbital radius of $\kappa_n = \sqrt{2 |n|}/\ell_B$. Due to the relative twist angle, the orbital center of the top layer is shifted by $\Delta K = 2 K_D \sin(2 \theta)$, where $K_D = | \mathbf{K}|$ is the position of the Dirac point of graphene. In this semiclassical picture, the tunneling amplitude between Landau levels is maximized when the fermion orbits touch each other, leading to the condition $\kappa_{n_\tfr} \pm \kappa_{n_\bfr} = \Delta K$. As the twist angle increases, the difference between the Landau levels must also increase to fulfill the condition. Increasing the magnetic field has the opposite effect since $\kappa_n \sim \sqrt{B}$. 

In Fig.~\ref{fig:magfield_el}b) we show the elastic tunneling current as a function of the applied bias voltage and the twist angle, assuming both graphene layers are at charge neutrality. Pronounced plateaus corresponding to transitions between different Landau levels can be seen. The width of the plateaus scales inversely with the magnetic length. Since the energy of Landau levels scales with $\eps_n \sim \sqrt{|n|}$, the separation between consecutive Landau levels decreases as $n$ increases. Consequently, only transitions between the first few Landau levels, where the separation $\Delta E = \eps_{n_\tfr}-\eps_{n_\bfr}$ is larger than the energy resolution, give a clean and unique signal to the tunneling current. 

\begin{figure}
\begin{center}
\includegraphics[width=\linewidth]{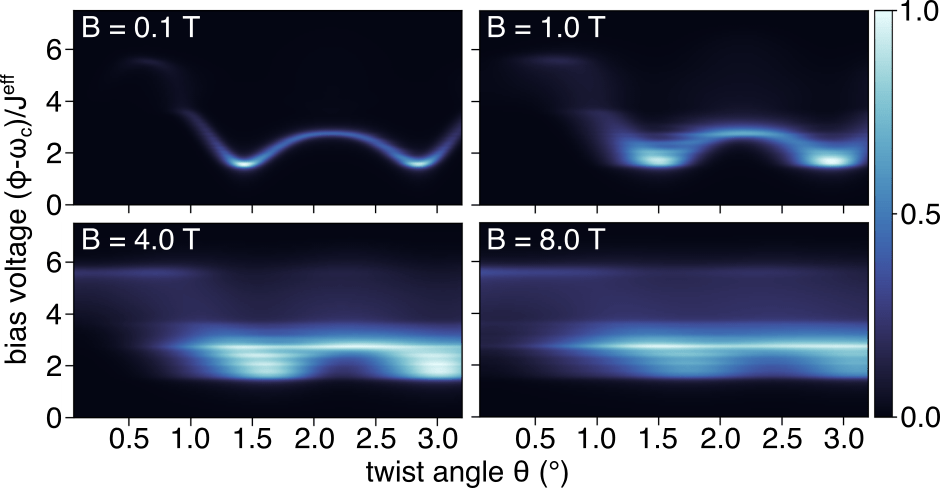}
\caption{\textbf{Inelastic tunneling in a magnetic field.}\textbf{a)} Inelastic contribution to the tunneling current for an intermediate layer in the CSL phase, measuring the dynamical spin-structure factor $\mathcal{S}(\kk, \omega)$. We show $\mathcal{S}_{01}(\omega)$ for different magnetic fields as a function of the twist angle. Due to the formation of Landau levels in the top and bottom graphene layer, momentum space resolution decreases with increasing external field strength. Since $\mathcal{S}_{01}(\omega)$ corresponds to a tunneling process from the zeroth to the first Landau level, we shift the bias voltage by $\omega_c$. In general, $\mathcal{S}_{nm}(\omega)$ gives contributions around a bias voltage of $(\sqrt{n}-\sqrt{m})\omega_c$. }
\label{fig:magfield}
\end{center}
\end{figure}

Information regarding collective excitations of the probe, sandwiched between the two graphene layers, is obtained from the inelastic contribution to the tunneling current, proportional to $|\Gamma_1|^2$. An analogous derivation to the one presented in Sec.~\ref{sec:inelastic} leads to an inelastic tunneling current given by 
\begin{align*}
	I^{(2)}(\phi) = &2 \pi e \Gamma_1^2 \displaystyle\sum_{n_\tfr n_\bfr} \big[ n_B(\varepsilon_{n_\tfr}-\varepsilon_{n_\bfr}+\phi)-n_B(\varepsilon_{n_\tfr}-\varepsilon_{n_\bfr})]\\ &\times [f(\varepsilon_{n_\tfr})-f(\varepsilon_{n_\bfr})]  \mathcal{S}_{n_\tfr n_\bfr} (\varepsilon_{n_\tfr}-\varepsilon_{n_\bfr}+\phi) \numberthis \label{eq:tun_curr_inel_mag}
\end{align*}
with
\begin{equation}
	\mathcal{S}_{n_\tfr n_\bfr}(\omega) = \frac{L^4}{(2\pi)^2 N} \int \dd k_\tfr \dd k_\bfr \displaystyle\sum_{\mathbf{k}} |\Psi^{n_\tfr k_\tfr, n_\bfr k_\bfr}_\mathbf{k}|^2 \mathcal{S}(\mathbf{k}, \omega)
\end{equation}
and
\begin{equation}
	\Psi^{n_\tfr k_\tfr, n_\bfr k_\bfr}_\mathbf{k} = \frac{1}{L^2} \int \text{d}^2 r \; \psi^*_{n_\tfr, k_\tfr}(\mathbf{r}) e^{i \mathbf{k} \cdot \mathbf{r}}  \psi_{n_\bfr, k_\bfr} (\mathbf{r}). \label{eq:coef_basis}
\end{equation}
Here, $n_\bfr(\eps)$ is the Bose-Einstein distribution, and $f(\eps)$ is the Fermi-Dirac distribution. Their only effect in Eq.~\eqref{eq:tun_curr_inel_mag} is ensuring that only energetically allowed transitions between an empty and a filled Landau level contribute to the current. Both top and bottom-layer graphene are held at charge neutrality. We see that the inelastic tunneling current still measures the dynamical structure factor 
\begin{equation}
    \mathcal{S}(\kk, \omega) = -i \displaystyle\int_0^\infty \dd t\; e^{i \omega t} \expval{[\hat{O}_\kk(t), \hat{O}_{-\kk}(0)]}
\end{equation}
but expressed in the Landau level basis. Explicit expressions for the coefficients of Eq.~\eqref{eq:coef_basis} are given in Appendix~\ref{app:4}. The twist angle dependence is implicitly contained in $\mathcal{S}_{n_\tfr n_\bfr}(\omega)$.
In Fig.~\ref{fig:magfield}, we show the inelastic contribution to the tunneling current for the transition between the zeroth and first Landau level for different magnetic fields, measuring the dynamical spin-structure factor of a CSL. Since the primary objective of this section is to characterize the response of the QTM in an external magnetic field, we assume a fixed $\mathcal{S}(\kk, \omega)$ independent of the magnetic field. 
The measured response $\mathcal{S}_{n_\tfr n_\bfr}(\omega)$ consists of a convolution of the structure factor $\mathcal{S}(\kk, \omega)$ and the tunneling matrix elements between the Landau levels $n_\tfr$ and $ n_\bfr$ of graphene. Even at a small magnetic field of $B=0.1$ T, the cyclotron frequency is already $\omega_c \approx 30$ meV, which is much larger than the typical energy scale of the effective spin-coupling $J^\text{eff}$ expected for the CSL. Despite the complex convolution, we can, therefore, easily reconstruct the inelastic scattering response of the system by focusing on bias voltages that are in the vicinity of the cyclotron frequency, which is the energy difference of the zeroth and the first Landau level.

As expected, one can see in Fig.~\ref{fig:magfield} that for small magnetic fields, the sharp collective mode of the CSL is still visible with good momentum resolution. As the magnetic field is increased, momentum resolution is lost. Nonetheless, even at a magnetic field of $B=8$ T, the signal is strongest at the $K$ and $K'$, a remnant of the fact that the collective mode has its minima there.
We conclude that the QTM can still provide useful momentum-resolved information about collective modes at finite magnetic fields. 

\begin{widetext}
\section{Expression for transformation to Landau-level basis} \label{app:4}

Here, we provide explicit expressions for Eq.~\eqref{eq:tunampl_el} and Eq.~\eqref{eq:coef_basis}. For that purpose, it is useful to consider the following integral
\begin{equation}
    I_\qb(n, m) := \displaystyle\int_{-\infty}^\infty \dd x\; e^{i q_x x} (\phi_n^{(\tfr) }(x))^* \phi_m^{(\bfr)}(x) \label{eq:int1}
\end{equation}
with $\phi_n^{(\ell)}(x)$ defined in Eq.~\eqref{eq:phistates}. To make progress, we first introduce another integral:
\begin{equation}
    F_{nm}(b, c) := \displaystyle\int_{-\infty}^\infty \dd z\; e^{ z^2 + b z -c^2/2} H_n(z) H_m(z-c).
\end{equation}

In terms of $F_{nm}$, the integral Eq.~\eqref{eq:int1} can be expressed as
\begin{equation}
    I_\qb(n, m) = \frac{e^{i q_x X_\tfr}}{\sqrt{2^{n+m}\pi n! m!}}F_{n m}\big(b=\ell_B[\Delta K_y-q_y + i(\Delta K_x + q_x)],\; c= \ell_B[\Delta K_y-q_y] \big)
\end{equation}
with $X_\ell = \ell_B^2 (k_y + \ell \Delta K_y)$ and $\Delta \mathbf{K} = (\mathrm{R}(\theta)-\mathbb{1})\mathbf{K}$. Here $\mathrm{R}(\theta)$ is a rotation matrix. Using standard properties of Hermite polynomials, $F_{nm}$ can be computed to be
\begin{equation}
    F_{nm}(b, c) = \sqrt{\pi} e^{b^2/4 -c^2/2} \displaystyle\sum_{j=0}^{m} \displaystyle\sum_{i=0}^{\text{min}(n,j)}(-c)^{m-j}b^{n+j-2i} \frac{2^{m+i-j}n!m!}{(m-j)!(n-i)!(j-i)!i!} 
\end{equation}
For Eq.~\eqref{eq:coef_basis}, we find
\begin{align*}
    \Psi^{n_\tfr k_\tfr, n_\bfr k_\bfr}_\qb = C_{n_\tfr} C_{n_\bfr} &\delta(k_\tfr-k_\bfr-q_y) \big[ I_\qb(|n_\tfr|, |n_\bfr|)
    + \text{sgn}(n_\tfr)\text{sgn}(n_\bfr) I_\qb(|n_\tfr|-1, |n_\bfr|-1) \\ &
    - i\; \text{sgn}(n_\bfr) I_\qb(|n_\tfr|, |n_\bfr|-1) + i\; \text{sgn}(n_\tfr) I_\qb(|n_\tfr|-1, |n_\bfr|)  \big] \numberthis
\end{align*}
with $C_n = \sqrt{(1+\delta_{n, 0})/2}$. For the tunneling amplitudes Eq.~\eqref{eq:tunampl_el} one obtains
\begin{equation}
    W(n_\tfr, n_\bfr) = \frac{L^4}{(2\pi)^2} \int \dd k_\tfr \dd k_\bfr \; |\Psi^{n_\tfr k_\tfr, n_\bfr k_\bfr}_{\qb=0}|^2. 
\end{equation}
    
\end{widetext}
\bibliography{bibfile}

\end{document}